\documentclass[aps, reprint,amsmath,amssymb,longbibliography]{revtex4-2}

\usepackage{graphicx} 
\usepackage{float}
\usepackage{adjustbox}
\usepackage{comment}
\usepackage{xcolor}
\usepackage{tikz}
\usepackage{tabularx}
\usepackage{makecell}
\usepackage{ragged2e }
\usepackage{lipsum}
\usepackage{capt-of}

\newcommand{\calA}{\mathcal{A}}
\newcommand{\calAzero}{\mathcal{A}_0}

\newcommand{\meancalA}{\langle \mathcal{A} \rangle}
\newcommand{\varcalA}{V_{\mathcal{A}}}
\newcommand{\skewcalA}{S_{\mathcal{A}}}

\begin{document}

\title{Cell Shape Emerges from Motion}

\author{Gautham Gopinath$^{1,2}$}
\thanks{These authors contributed equally.}
\author{Emmanuel Y. Mintah$^{3,2}$}
\thanks{These authors contributed equally.}
\author{Aashrith Saraswathibhatla$^4$}
\author{Jonah J. Spencer$^{5,6}$}
\author{Shahar Nahum$^7$}
\author{Lior Atia$^7$}
\author{Jacob Notbohm$^{5,6}$}
\author{Mark D. Shattuck$^8$}
\author{Corey S. O'Hern$^{9,1,10,11,2}$}

	 \affiliation{
	 	$^1$Department of Physics, Yale University, New Haven, Connecticut 06520, USA \\   
        $^2$Integrated Graduate Program in Physical and Engineering Biology, Yale University, New Haven, Connecticut 06520, USA\\ 
	 	$^3$Department of Biomedical Engineering, Yale University, New Haven, Connecticut 06520, USA\\
        $^4$Department of Biomedical Engineering, University of Minnesota, Minneapolis, Minnesota 55455, USA \\
        $^5$Biophysics Program, University of Wisconsin-Madison, 1500 Engineering Drive, Madison,
        Wisconsin 53706, USA\\
        $^6$Department of Mechanical Engineering, University of Wisconsin-Madison, 1500 Engineering Drive, Madison,
        Wisconsin 53706, USA\\
        $^7$Department of Mechanical Engineering, Ben Gurion University of the Negev, Be'er Sheve, Israel\\
         $^8$Benjamin Levich Institute and Physics Department, City College of New York, New York, New York 10031, USA\\
	 	$^9$Department of Mechanical Engineering, Yale University, New Haven, Connecticut 06520, USA\\
       $^{10}$Department of Applied Physics, Yale University, New Haven, Connecticut 06520, USA\\
         $^{11}$Graduate Program in Computational Biology and Biomedical Informatics, Yale University, New Haven, Connecticut 06520, USA
         }
\begin{abstract}
We perform cell segmentation on images from experimental studies of confluent, mobile cells in epithelial monolayers and show that these systems possess a broad, positively-skewed shape parameter distribution $P({\cal A})$, where ${\cal A}=p^2/4\pi a$, $p$ is the perimeter, and $a$ is area of each cell. $P({\cal A})$ is peaked at a value higher than the typical shape parameter ${\cal A}^* \sim 1.15$ that occurs for randomly packed, static confluent cell monolayers.  The distribution does not arise from a heterogeneous population of cells with different {\it fixed} ${\cal A}$, nor can it arise from cell shape fluctuations from strains below the elastic limit. Instead, we find that all cells in each monolayer sample ${\cal A}$ values that span the full shape parameter distribution. We develop a deformable particle model that allows cell perimeter to adapt to local forces during cell motion, and this model recovers $P({\cal A})$ to within $5\%$ for both MDCK and HaCaT epithelial cell monolayers. These results emphasize that confluent epithelial monolayers of mobile cells generate a well-defined broad shape parameter distribution that is independent of the initial cell shapes.

\end{abstract}

\maketitle

\section{Introduction}
Packings of rigid particles, such as colloidal glasses \cite{pusey1986phase} and jammed granular materials \cite{behringer2018physics}, form solid-like states at high packing fractions $\phi$ where the particles are unable to rearrange at long timescales since there is insufficient free volume and the particles have fixed shapes \cite{PhysRevLett.59.2083}. In contrast, epithelial cells in confluent monolayers with $\phi=1.0$ can be highly mobile since each cell can undergo dynamic shape changes~\cite{doostmohammadi2015celebrating, saraswathibhatla2020tractions}. The ability of epithelial cells to adjust their shapes to enable motion raises several important questions. Do epithelial cells select a preferred set of shapes and their motion results from these preselected shapes, or are the cell shapes an output of the cell motion? Further, by analyzing the cell shape distributions from mobile, confluent epithelial monolayers, can we determine whether cell shapes deform reversibly or irreversibly?

Previous experimental studies of confluent epithelial cell monolayers have performed Voronoi tessellations of the cell nuclei and shown that the cell shape parameter distributions are broad \cite{atia2018geometric} with significant weight above the typical minimum value for static monolayers~\cite{boromand2018jamming}. One explanation is that all cells have the same preferred shape, but the cells undergo significant elastic deformations during motion. However, previous experimental studies have shown that even $10\%$ strain can cause plastic deformation in the cell cortex~\cite{fernandez2008single,cordes2020prestress, chugh2018actin}. Another explanation for this broad shape parameter distribution is that epithelial cell monolayers are heterogeneous, composed of elastic cells that each possess different preferred shape parameters around which they fluctuate. Thus, in many prior modeling studies, cell shape is viewed as an input to the system from which the dynamics and mechanical properties of the system emerge~\cite{bi2015density, bi2016voronoi,merkel2018geometrically}. A shortcoming of this description is that it does not explain why epithelial monolayers with different cell types have similar cell shape parameter distributions~\cite{atia2018geometric, sadhukhan2022origin}. In this work, we propose that well-defined broad cell shape parameter distributions emerge in confluent, mobile epithelial cell monolayers because the perimeters of individual cells can deform irreversibly, which allows the cells to explore a wide range of shapes while moving in confluent monolayers.  

Several of us have performed prior experimental studies to measure cell shapes in confluent epithelial cell monolayers from different cell lines and over a range of cell densities. These prior studies include time-lapse phase contrast imaging of (1) low- and (2) high-density islands of MDCK cells \cite{saraswathibhatla2020spatiotemporal,saraswathibhatla2020tractions}, (3) low-density islands of human keratinocyte (HaCaT) cells, (4) time-lapse phase contrast and GFP-tagged migratory MDCK epithelial cell monolayers, and (5) asthmatic and (6) non-asthmatic human bronchial (HB) epithelial cells that have been chemically fixed at different air-liquid interface culturing times \cite{atia2018geometric}. (See Appendix A for experimental details.) We characterize cell monolayers as low- (high-) density when their number density $\rho < \rho_0$ ($\rho > \rho_0$), where $\rho_0 \sim 4$,$000$-$7$,$000$ cells/mm$^2$.  Above $\rho_0$, MDCK cell monolayers experience density-related growth inhibition~\cite{rosen1980cell} and HaCaT cells form partial multilayers \cite{szabo2001cell}. Similarly, even at densities below $\rho_0$, HB cell monolayers can exhibit decreased cell motility due to matured cell-cell and cell-substrate junctions \cite{garcia2015physics}. We perform image analyses of these six datasets using Cellpose \cite{stringer2021cellpose, pachitariu2022cellpose} augmented by manual tracing, which allows us to identify curved cell boundaries in contrast to Voronoi tessellation of the cell nuclei \cite{atia2018geometric}. We then quantify the cell shapes by calculating the shape parameter $\calA=p^2/4\pi a$, where $p$ and $a$ are the perimeter and area of each cell, respectively. We find that highly mobile confluent epithelial cell monolayers exhibit a well-defined broad and positively-skewed shape parameter distribution $P({\cal A})$ with mean shape parameters greater than the minimum ${\cal A}_{\rm min} \sim 1.15$ that typically occurs for static confluent cell monolayers~\cite{boromand2018jamming}. Each cell spans a wide range of shape parameters during motion, and the time-averaged distribution of shape parameters is similar to the spatially averaged distribution. To understand the shape parameter distribution, we develop a deformable particle model (DPM) with an adaptive preferred perimeter that can vary based on local forces during cell motion. The DPM captures the shape parameter distribution of mobile epithelial cell monolayers to within $5\%$ for both MDCK and HaCaT cells. These results emphasize that the cell shape parameter distribution is not generated by small fluctuations around a fixed cell shape parameter. Instead, the cell shape parameter distribution is independent of the initial cell shapes and reaches a broad, steady-state cell shape distribution that is relatively insensitive to the cell mechanical properties.  

This article is organized into three remaining sections. Sec.~\ref{methods_section} focuses on the experimental and computational methods used in our studies and is divided into four subsections. The first subsection describes the cell segmentation pipeline. The next two subsections introduce two computational models for modeling mobile epithelial cells in confluent monolayers: (1) the deformable particle model (DPM) with a harmonic energy penalty for deviations in the cell perimeter from its preferred value~\cite{atia2018geometric,bi2015density, bi2016voronoi, merkel2018geometrically} and (2) the DPM except with an adaptive preferred perimeter. In the final subsection, we describe the values of the model parameters that we choose. In Sec.~\ref{results_section}, we describe the main results, including the characterization of the cell shape parameter distributions obtained from the experimental studies and the predicted cell shape parameter distributions from the two computational models. In Sec.~\ref{discussion_section},  we discuss the possible biological mechanisms that can give rise to irreversible changes in cell perimeter, including the folding and unfolding of membrane reservoirs \cite{gervasio2011caveolae, fernandez2008single} and membrane endocytosis and exocytosis \cite{apodaca2002modulation, sheetz2001cell}. Cells can also change their shapes via area fluctuations generated by fluid flows through gap junctions~\cite{zehnder2015cell} and in this section we comment on how cell shapes can be influenced by changes in the preferred area. 
We also describe several interesting future directions, for example, identifying the key parameters that control when cells in confluent monolayers are solid- versus fluid-like and the connection between the rate of cell neighbor exchanges and the cell shape parameter distribution.

\section{Methods}
\label{methods_section}

The methods section contains four subsections. In Sec.~\ref{image_analysis}, we describe the methods used to segment the cells, measure their perimeter and area, calculate the cell shape parameter ${\cal A}$, and track the centers of mass of mobile epithelial cells as a function of time. In Sec.~\ref{sec:dpm_methods_fixed}, we introduce the deformable particle model (DPM) with a harmonic energy penalty for deviations in the cell perimeter from a preferred value and simulations of the DPM with active Brownian force $f_0$ and persistence time $\tau$. In Sec.~\ref{dpm_methods_plastic}, we describe another model for mobile epithelial cells, the active Brownian DPM with an adaptive preferred perimeter, where the preferred perimeter can change over a time scale $\tau_p$ in response to local forces during cell motion. Finally in Sec.~\ref{sec:simulation_parameters}, we describe the simulation parameters that we use for these two computational models. We fix four of the relevant dimensionless parameters based on biophysical constraints. We then vary a combination of the dimensionless active Brownian force $\widetilde{f}_0$ and persistence time $\widetilde{\tau}$, and the perimeter relaxation time scale $\widetilde{\tau}_p$ to investigate how these variables affect the emergent cell shape parameter distributions.

\subsection{Image Segmentation and Analysis}
\label{image_analysis}

\begin{figure}[tb]
    \includegraphics[width=\columnwidth]{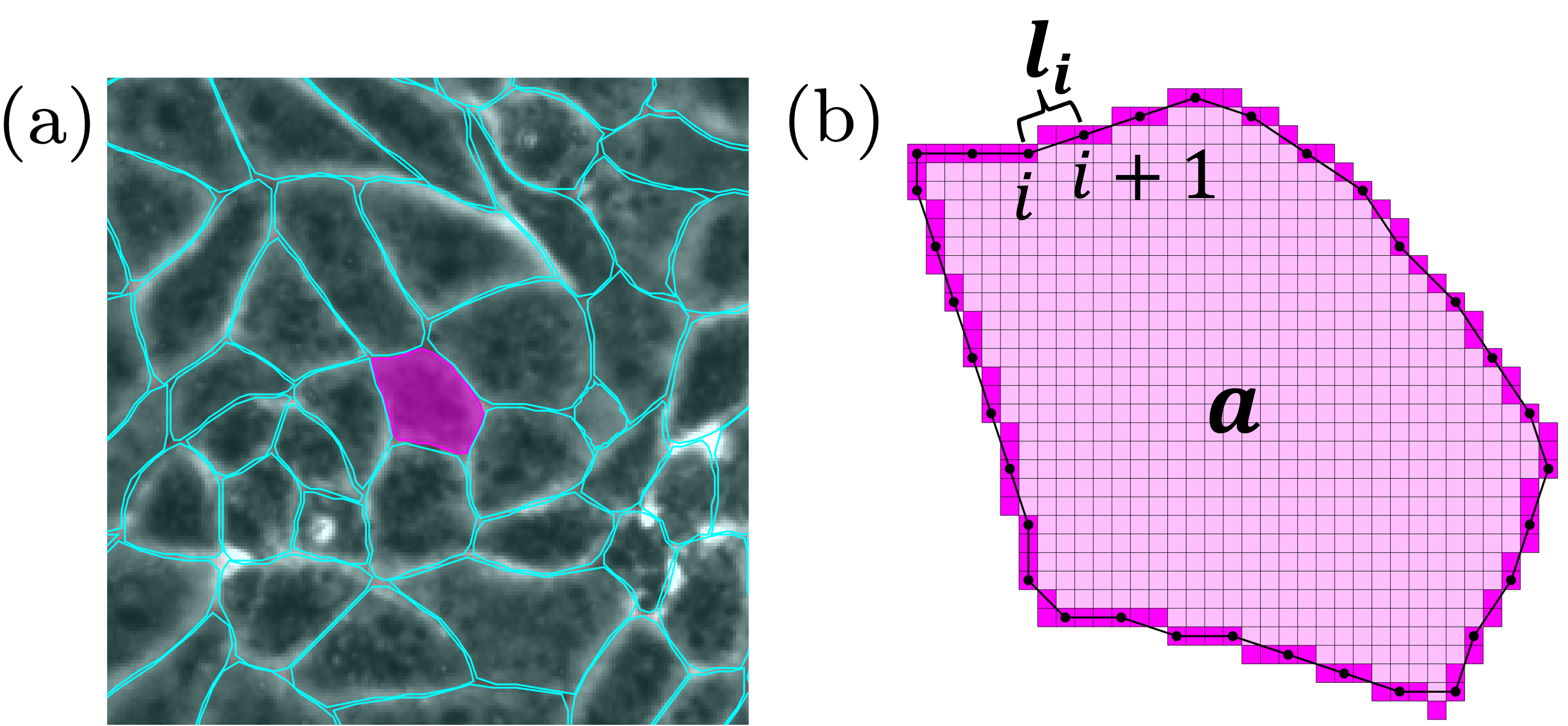}
    \caption{(a) A segmented section of a phase contrast image of a low-density island of MDCK epithelial cells. The cyan borders mark the cell boundaries. (b) The highlighted magenta cell in (a), showing the pixels that make up the cell and the $N_v$ boundary vertices (black points) that are used to calculate the cell shape parameter $\calA= p^2 / 4\pi a \approx 1.22$, where $p$ is the perimeter, $a$ is the area, and $l_i$ is the distance between consecutive vertices $i$ and $i+1$.
    }
    \label{fig:segment}
\end{figure}

To determine the shapes of the cells from the experimental images of the epithelial cell monolayers, we first pass the images to Cellpose \cite{stringer2021cellpose, pachitariu2022cellpose} and use one of the built-in segmentation algorithms (e.g. cyto2) to obtain a baseline segmentation of the monolayer in two dimensions, i.e. a list of pixels (or mask) that is assigned to each cell. We then manually correct the segmentation by identifying cells missed by Cellpose and removing dead cells that were identified by Cellpose. (See Fig.~\ref{fig:segment} (a).) The boundary pixels of each cell mask are filtered so that each mask has $N_v \sim 25$ evenly spaced vertices located at the centers of the boundary pixels, ${\vec r}_i=\left(x_i,y_i \right)$ as shown in Fig. \ref{fig:segment} (b). We calculate the perimeter $p=\sum_{i=1}^{N_v}l_i$, where  $l_i=|\vec{r}_{i+1}-\vec{r}_i|$, and the area $a=\frac{1}{2}\sum_{i=1}^{N_v}\left(x_iy_{i+1} - x_{i+1}y_i \right)$ to determine the shape parameter $\calA =p^2/(4 \pi a)$ of each cell. (We find that the error from calculations of the cell shape parameter at the pixel resolution from the images of MDCK and HaCaT islands is nearly constant when $20 \leq N_v \leq 40$.) 

\begin{figure}[tb]
    \includegraphics[width=\columnwidth]{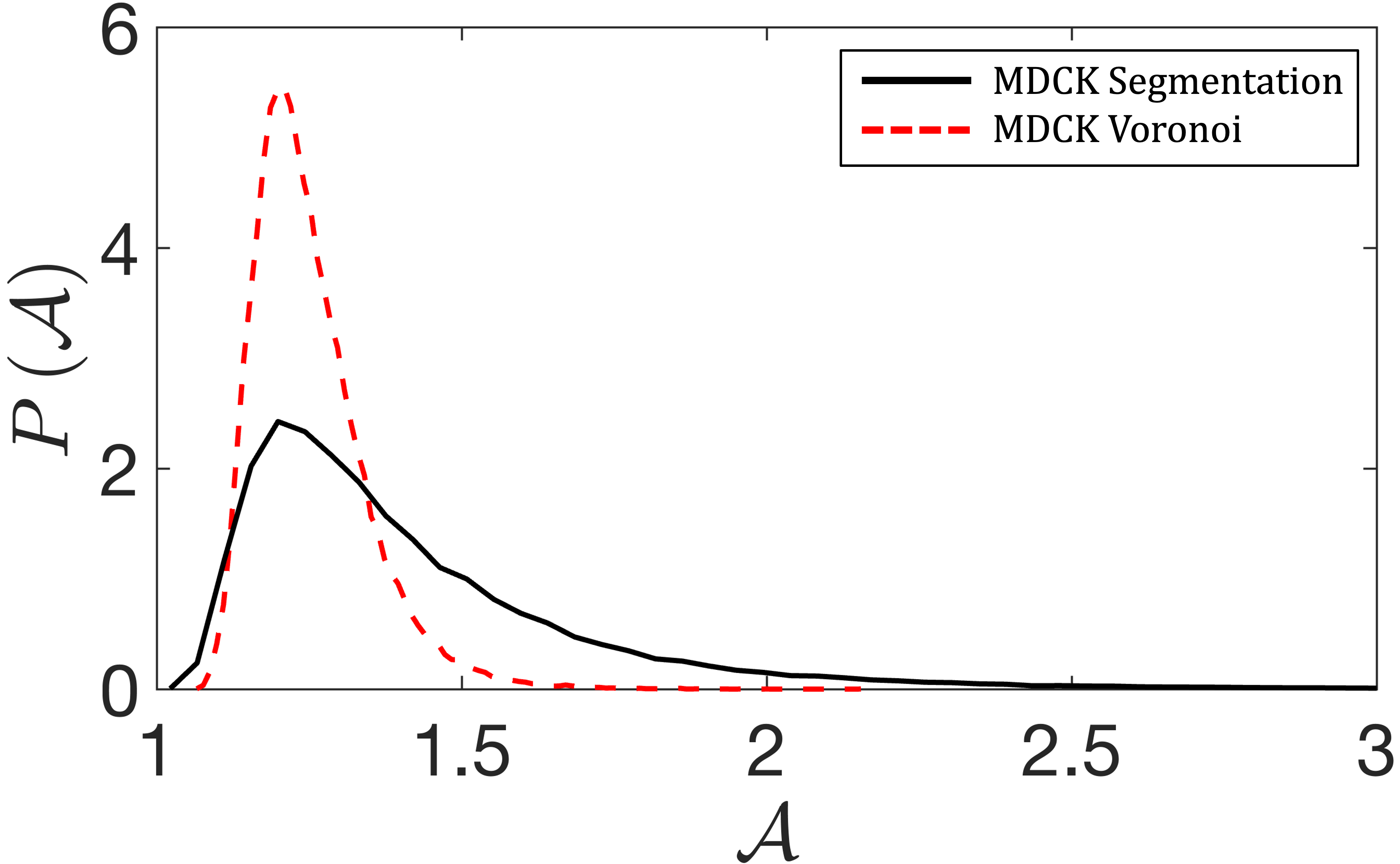}
    
    \caption{Probability distributions of the cell shape parameter $P(\cal{A})$ from the segmentation of low-density islands of MDCK epithelial cells using Cellpose (black solid) and Voronoi tessellations of the centers of mass of the cell masks from the segmentation (red dashed).
    }
    \label{fig:voronoi}
\end{figure}

Previous studies have characterized the sizes and shapes of epithelial cells through Voronoi tessellations of the cell nuclei \cite{atia2018geometric, roshal2022random} and assigned the cell area and perimeter to be the area and perimeter of the corresponding Voronoi polygon that encloses the cell nucleus. However, Voronoi tessellation enforces straight edges between cells and therefore cannot capture the curvature of epithelial cell boundaries. In Fig.~\ref{fig:voronoi}, we compare the cell shape parameter distributions $P(\cal{A})$ from cell segmentation using Cellpose (black solid line) of the low-density islands of MDCK epithelial cells to that from Voronoi tessellations of the centers of mass of the cell masks (red dashed line). We find that the true distribution of MDCK epithelial cell shape parameters $P({\cal A})$ is broad and positively-skewed with $\langle A \rangle \sim 1.4$, while $P({\cal A})$ from the Voronoi tessellation is much narrower and peaked near $\langle A \rangle \sim 1.2$. Similarly, we find that the true distributions $P({\cal A})$ from the cell segmentations of the other five datasets are broader than the corresponding distributions obtained from Voronoi tessellation. Thus, using Voronoi tessellations does not accurately capture the cell shape parameter distribution of epithelial cell monolayers.  

\begin{figure*}[htbp]
    \centering
    \includegraphics[width=\textwidth]{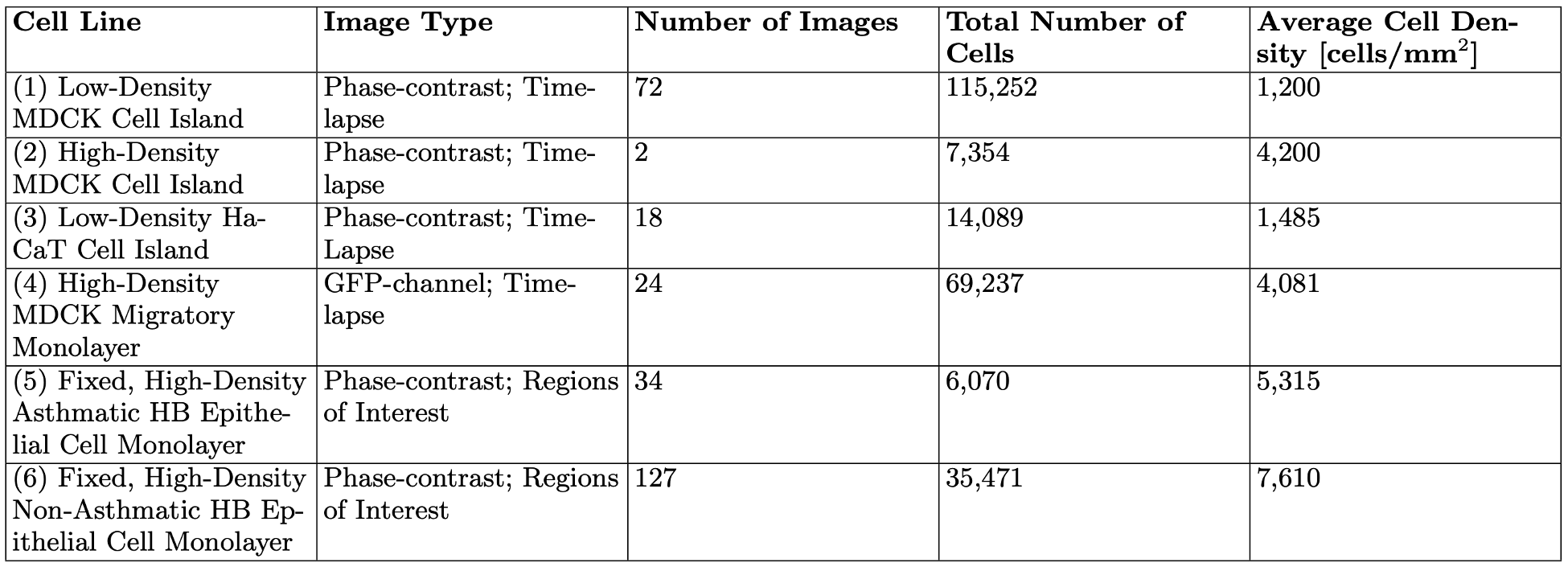}   
    
    \captionof{table}{Descriptions of the six datasets of epithelial cells including the cell line, imaging technique, number of images, total number of segmented cells, and average cell density of the monolayer.}
    \label{table:image_details}
\end{figure*}

The six datasets of images containing epithelial cell monolayers are summarized in Table~\ref{table:image_details}. Most of the images were obtained from phase-contrast microscopy with the exception of the GFP-labeled MDCK cells. From these datasets, we segmented nearly $250$,$000$ cells, about half of which were from the time-lapse images (over three hours) of the low-density MDCK epithelial cell islands. With the time-lapse images, we can track the centers of mass and shapes of individual cells within the monolayers over time. Individual cell tracking is performed by connecting the cell masks in the segmentation of an image at a given time with the nearest centers of mass of the cell masks in the image at the next time point. We compare the distribution of cell shape parameters from individual cells over time to the distributions of all cells in a given system at fixed times. The rates of apoptosis in the MDCK epithelial cell islands are small, i.e. $< 1\%$.  When a cell dies, it is removed from the cell tracking data set both before and after its death.  Mitotic divisions are treated as the addition of new daughter cells in the monolayer at the time of the division. We did not use the linear assignment problem for cell tracking since the cells undergo large shape changes between time points~\cite{ershov2022trackmate,tinevez2017trackmate}. 

\subsection{Model 1: Deformable Particle Model with Fixed Preferred Shape Parameter}

\label{sec:dpm_methods_fixed}

\begin{figure}[h]

\includegraphics[width=\columnwidth]{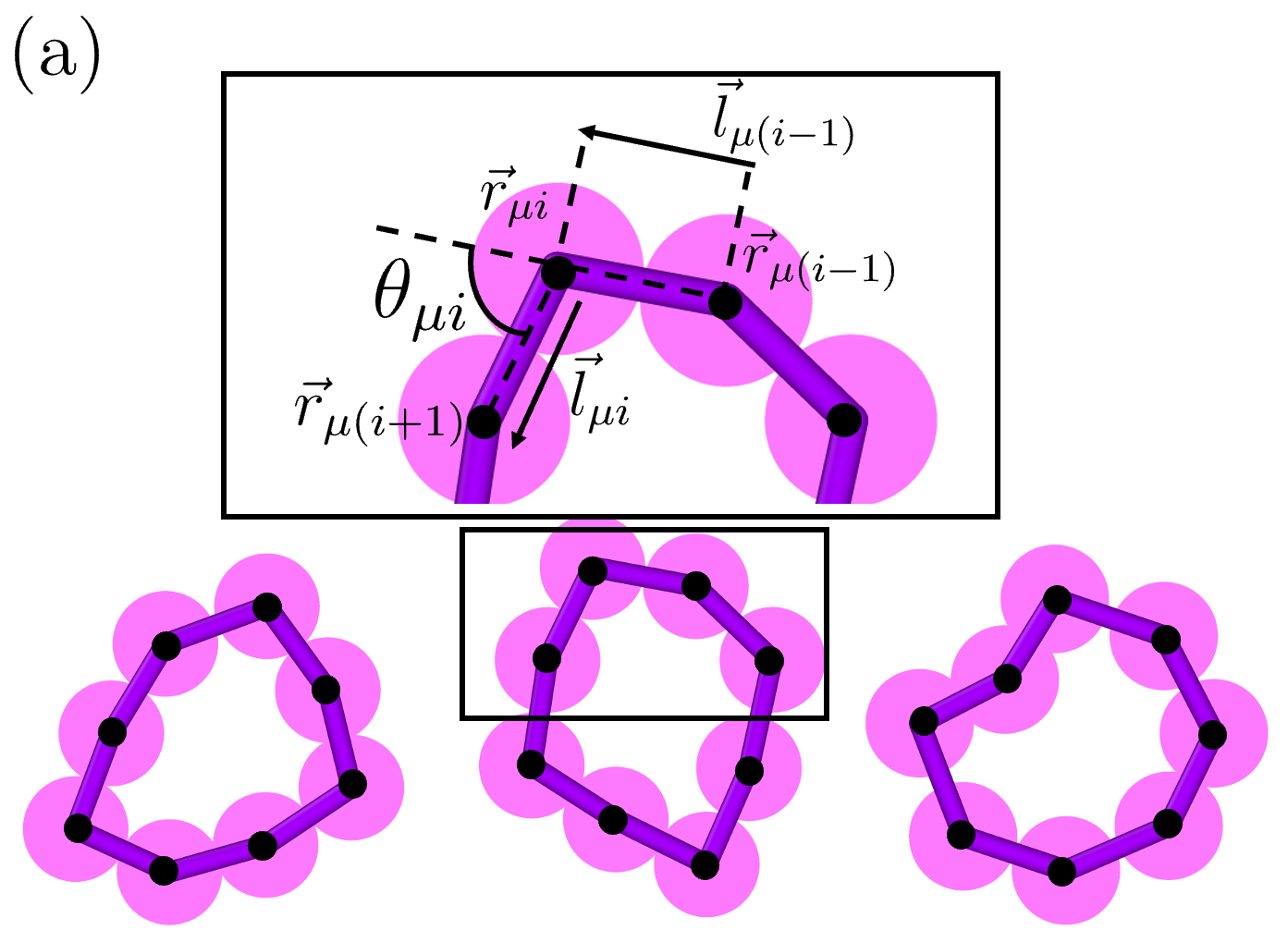}
\includegraphics[width =\columnwidth]{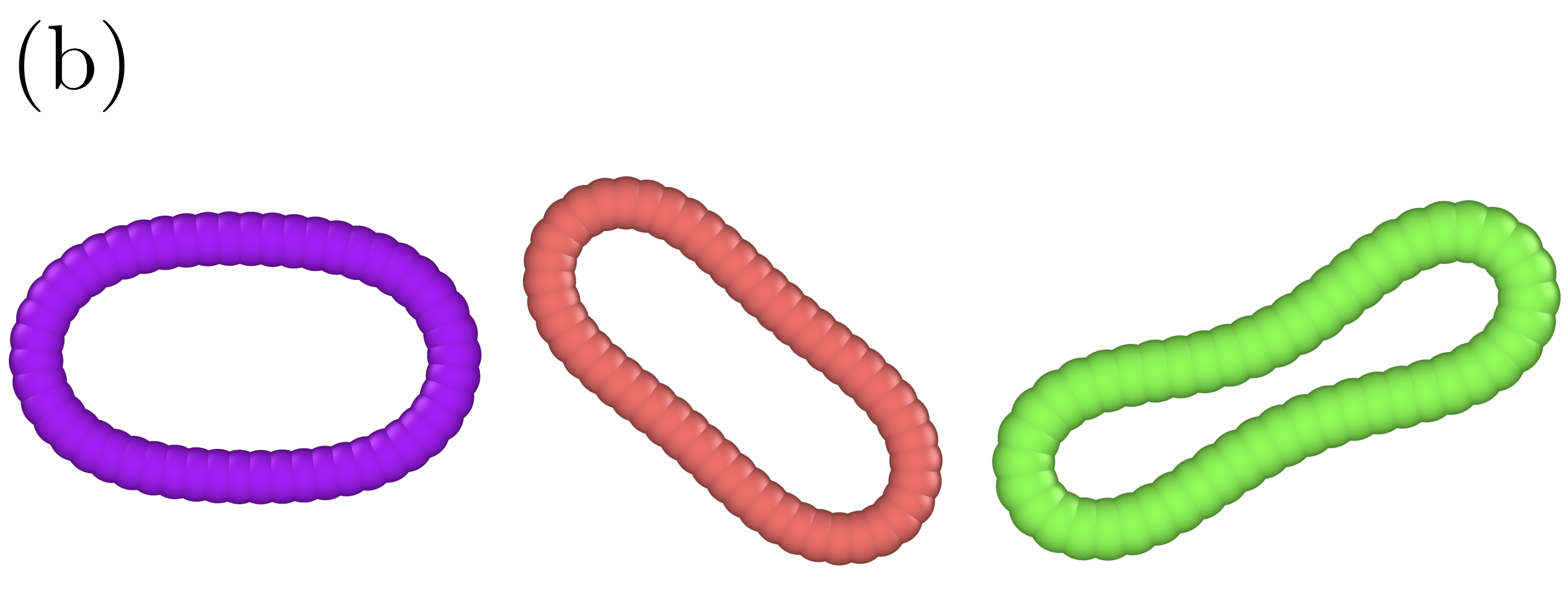}
\caption{(a) Three example shapes for a deformable particle with $U_{\mu,{\rm shape},0} \approx 0$ with preferred shape parameter $\calAzero=1.15$, $N_v=8$ vertices, and bending energy $\epsilon_b=0$. The rightmost shape has an invagination. The inset describes the geometry of the particle surface, where $\vec{r}_{\mu (i-1)}$, $\vec{r}_{\mu i}$, and $\vec{r}_{\mu (i+1)}$ are the positions of the vertices $i-1$, $i$, and $i+1$ and $\theta_{\mu i}$ is the angle between the bond vectors ${\hat l}_{\mu i}$ and ${\hat l}_{\mu(i-1)}$. (b) The minimum energy configurations for single deformable particles with $\epsilon_b/(k_l\sigma^2)=10^{-3}$, where $\sigma$ is the vertex diameter, $N_v=50$, and $\calAzero=1.15$, $1.4$, and $2$ from left to right.
    }
\label{fig:shape_ensemble}
\end{figure}

The deformable particle model (DPM) can describe the shapes of  cells by modeling each cell as a polygon with a discrete set of $N_v$ vertices on its surface \cite{boromand2018jamming}. Typically, $N_v \sim 50$ is large enough to capture the curved boundaries of epithelial cells. Given the coordinates of the vertices ${\vec r}_{\mu i}$ on polygon $\mu$, as shown in Fig. \ref{fig:shape_ensemble} (a), we can calculate the perimeter and area of the polygon, which determines its shape parameter $\calA$. 

We can constrain the shape parameter of deformable particle $\mu$ using the shape-energy function: 
\begin{equation}
\label{eq:shape_energy}
        U_{\mu,\text{shape}, 0} = \frac{\epsilon_l}{2 N_v}\sum_{i=1}^{N_v}\left(\frac{l_{\mu i}}{l_{\mu 0}} -1\right)^2  + \frac{\epsilon_a}{2}\left(\frac{a_\mu}{a_{\mu 0}} -1\right)^2,
\end{equation}
where $l_{\mu i}$ is the distance between adjacent vertices in polygon $\mu$ that are connected by a spring of rest length $l_{\mu0}$. (See Fig. \ref{fig:shape_ensemble} (a).) $a_\mu$ and $a_{\mu0}$ represent the area and preferred area of cell $\mu$, respectively. The area constraint with strength $\epsilon_a/a_{\mu0}^2$ represents the incompressibility of the fluid inside the cell. The preferred perimeter is $p_{\mu0} = N_v l_{\mu0}$, and the spring constant $k_l = \epsilon_l/l_{\mu0}^2$ controls the fluctuations about this preferred value.  Thus, the preferred shape parameter is $\calAzero = p_{\mu0}^2/(4\pi a_{\mu0})$.

One hypothesis for how cells sample their shape space is that cells sample shapes at fixed shape parameter $\calA_0$. Even at fixed preferred shape parameter, a cell can possess an ensemble of different shapes, each given by the positions of the vertices of the polygon.  We depict a few example shapes out of the ensemble of configurations at fixed shape parameter $\calA_0=1.15$ for a deformable particle with $N_v=8$ vertices in Fig.~\ref{fig:shape_ensemble} (a). All of these configurations have $U_{\mu,\text{shape},0} \approx 0$ and can possess invaginations, like that for the rightmost polygon in Fig.~\ref{fig:shape_ensemble} (a). We do not observe such concave shapes in epithelial cell monolayers, which motivates us to add an additional bending energy cost~\cite{nagle2015true}.

To penalize concave DPM shapes, we add a bending energy to the shape-energy function \cite{treado2021bridging}:
\begin{equation}
\label{eq:shape_energy}
        U_{\mu,\text{shape}} = U_{\mu,\text{shape},0}+
        \frac{\epsilon_b N_v}{2}\sum_{i=1}^{N_v}\theta_{\mu i}^2, 
\end{equation}
where $\theta_{\mu i}$ is the angle between ${\hat l}_{\mu i}= {\vec l}_{\mu i}/l_{\mu i}$ and ${\hat l}_{\mu(i+1)}$, and $\vec{l}_{\mu i}$ connects vertex $i$ to $i+1$ as shown in Fig. \ref{fig:shape_ensemble} (a).  
Setting $\epsilon_b > 0$ results in a single minimum energy conformation with $U_{\mu,\text{shape}} >0$ for a deformable particle at each $\calAzero$, in contrast to an ensemble of shapes with $U_{\mu,\text{shape}, 0} \approx 0$. The ground state for $U_{\mu,\text{shape}}$ is a pill-shaped DPM for $\calAzero < \calAzero^*(\epsilon_b)$, which scales with $\epsilon_b$, whereas the dumbbell shape is the ground state for $\calAzero > \calAzero^*(\epsilon_b)$ as shown in Fig.~ \ref{fig:shape_ensemble} (b) \cite{treado2021bridging}.

\begin{figure}[tb]
    \includegraphics[width=\columnwidth]{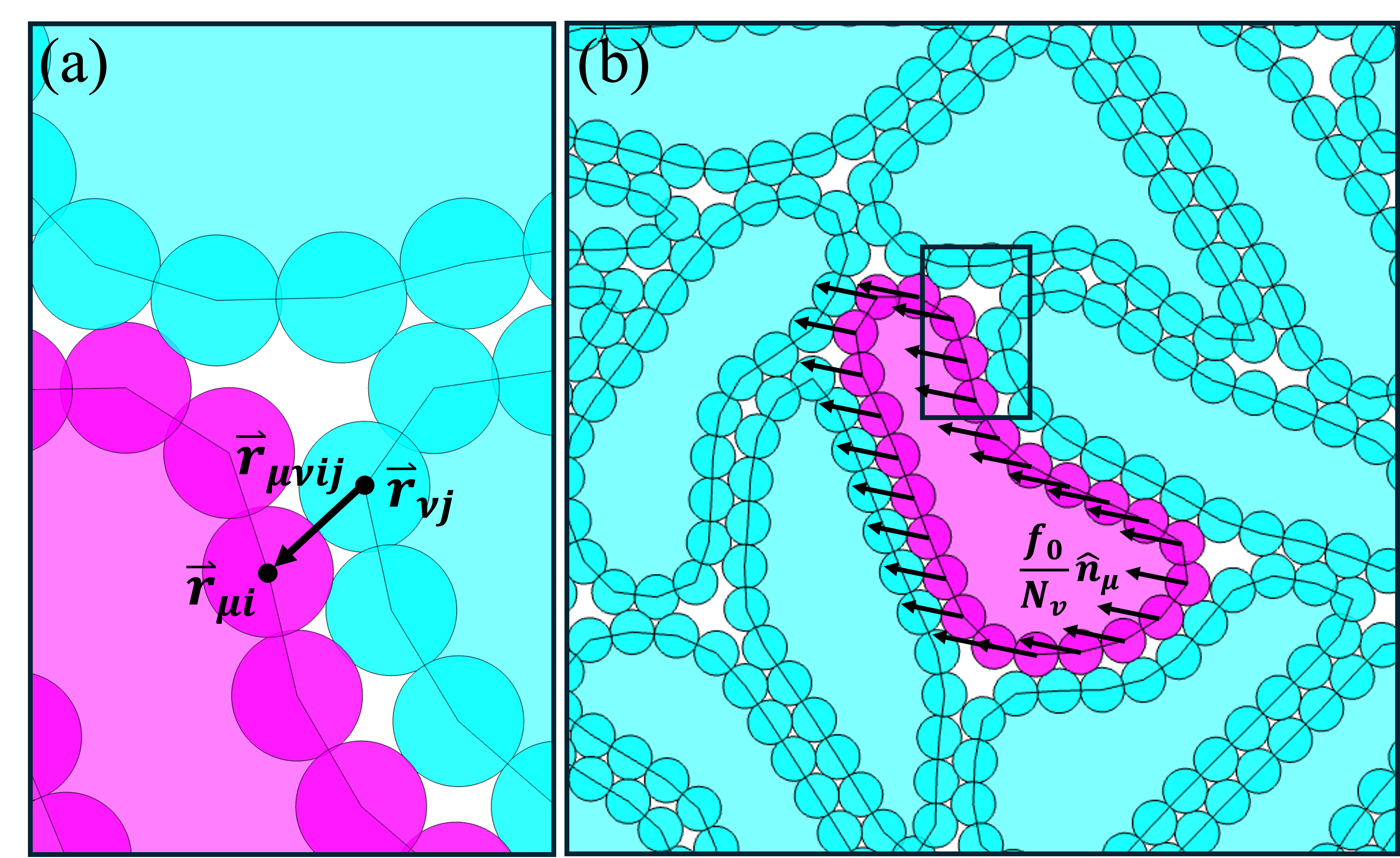}
    \includegraphics[width=\columnwidth]{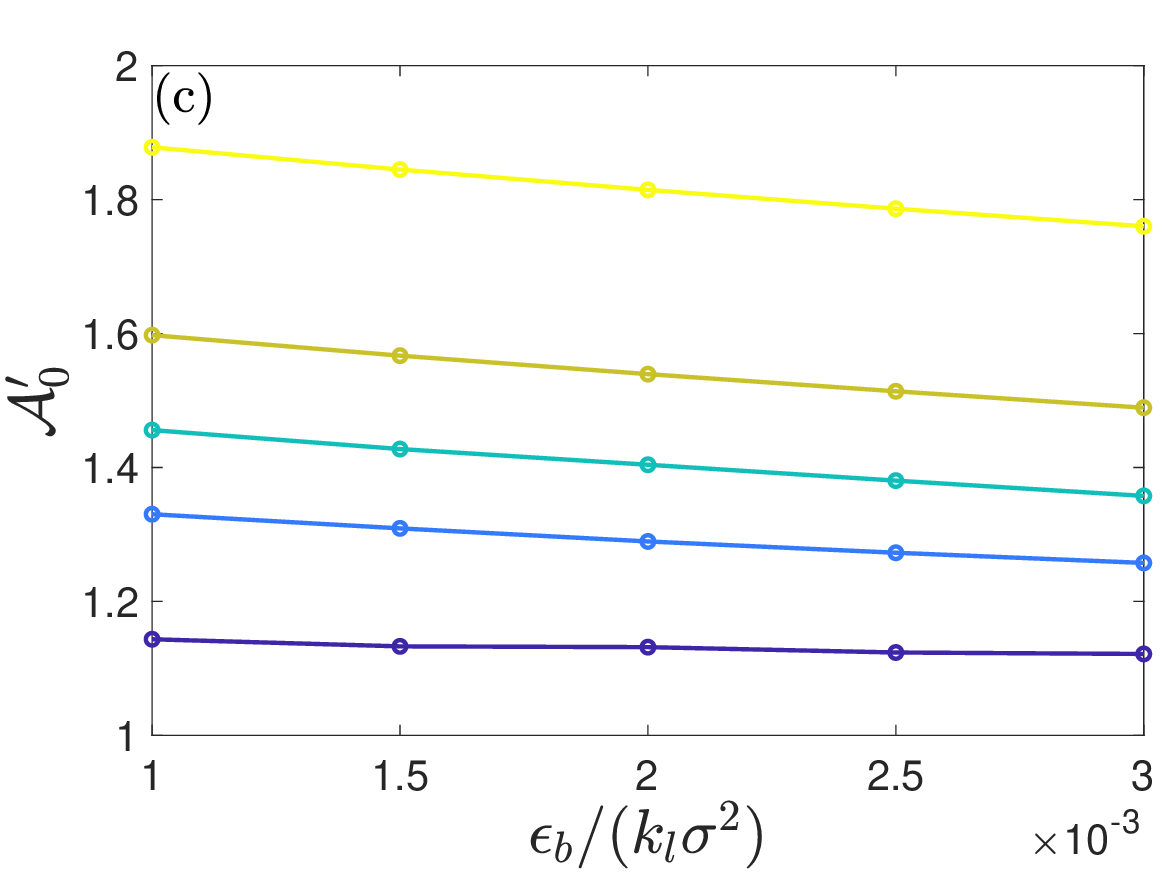}
        \caption{(a) A close-up of the magenta cell $\mu$ in (b). The separation vector between two vertices on different cells is $\vec{r}_{\mu \nu ij} = \vec{r}_{\mu i} - \vec{r}_{\nu j}$, where $\vec{r}_{\mu i}$ is the position of vertex $i$ on cell $\mu$ (magenta) and $\vec{r}_{\nu j}$ is the position of vertex $j$ on cell $\nu$ (cyan). (b) Schematic of a mechanically stable packing of deformable particles, each with $N_{v}=25$ vertices. The arrows indicate the active force $\left( f_{0}/N_{v} \right)\hat{n}_{\mu}$ that is applied to each vertex of cell $\mu$. (c) The mean shape parameter of mechanical stable packings $\calAzero'$ plotted as a function of the dimensionless bending energy $\epsilon_b/(k_l\sigma^2)$ for $\calAzero=1.15$, $1.4$, $1.55$, $1.7$, and $2$ from bottom to top. 
    }
    \label{fig:DPM}
\end{figure}

We assume that neighboring deformable particles $\mu$ and $\nu$ interact via purely repulsive, pairwise forces between overlapping vertices:
\begin{equation}
       U^{\mu\nu}_{\rm int} = \frac{\epsilon_v}{2}
     \sum_{i=1}^{N_v}\sum_{j=1}^{N_v}
     \left(1-\frac{r_{\mu\nu ij}}{\sigma} \right)^{2}
    \Theta\left(1-\frac{r_{\mu\nu ij}}{\sigma}\right),
\end{equation}
where $r_{\mu\nu ij}$ is the distance between vertex $i$ (with diameter $\sigma=2.5 l_{\mu 0}$) on cell $\mu$ and vertex $j$ on cell $\nu$ as shown in Fig.~\ref{fig:DPM} (a). The spring constant $\epsilon_v/\sigma^2$ controls the strength of the repulsion between overlapping vertices, and the Heaviside step function $\Theta$ ensures that the vertices only interact when they overlap. The total energy for a system of $N$ deformable particles with fixed preferred shape parameter is $U=\sum_{\mu=1}^N \left( U_{\mu, \text{shape}}+\sum_{\nu>\mu}U^{\mu \nu}_{\rm int} \right)$.

To model confluent monolayers of mobile cells, we begin by generating mechanically stable packings of deformable particles at packing fraction $\phi=0.95$.  We will then induce cell motion within these packings via active Brownian forces applied to each cell. To generate the cell packings, we initialize $N=50$ deformable particles at $\calAzero$ and $\phi=0.4$ in a periodic square simulation box with no overlaps between vertices on different cells. We then increase the packing fraction in steps of $\Delta\phi=10^{-4}$, relaxing the system after each step by integrating the following equation of motion:
\begin{equation}
\label{eq:EOM_packing}
    m\frac{d^2\vec{r}_{\mu i}}{dt^2} =  -\overrightarrow{\nabla}_{r_{\mu i}} U-\beta \frac{d\vec{r}_{\mu i}}{dt},
\end{equation}
 where $m$ is the vertex mass, $\vec{r}_{\mu i}$ is the vertex position, and $\beta/\sqrt{mk_l}=0.4$ is the dimensionless viscous drag such that the system is in the overdamped regime. We integrate Eq.~\ref{eq:EOM_packing} (and all other equations of motion) using a generalization of the BBK algorithm \cite{brunger1984stochastic}, which is second-order accurate in $\Delta t$~\cite{vanden2006second}, using timestep $\Delta t/\sqrt{m/k_l}=10^{-2}$.  After each increase in packing fraction,  we relax the system until $[U(N_t +1)-U(N_t)]/U(N_t) < 10^{-4}$, where $N_t$ is the total number of time steps, and continue this process until $\phi=0.95$.

After obtaining a mechanically stable packing of deformable particles at $\phi=0.95$, we drive the cells by adding active Brownian forces with equal magnitude $f_0/N_v$ to each vertex. (See Fig. \ref{fig:DPM} (b).) The equation of motion for each vertex is then:
\begin{equation}
\label{eq:EOM}
    m\frac{d^2\vec{r}_{\mu i}}{dt^2} =  -\overrightarrow{\nabla}_{r_{\mu i}} U+\frac{f_0}{N_v}\hat{n}_{\mu}-\beta \frac{d\vec{r}_{\mu i}}{dt},
\end{equation}
where the direction $\hat{n}_\mu=(\cos\alpha_\mu, \sin\alpha_\mu)$ and $\alpha_\mu$ is the angle that the active force makes with the $x$-axis. $\alpha_{\mu}$ evolves randomly according to $d\alpha_\mu/dt = \sqrt{2/(\tau\Delta t)} w_\mu$, where $w_\mu$ is a Gaussian random variable with mean zero and unit variance. As $\tau \to 0$, a randomly oriented force is applied at every timestep. In contrast, as $\tau \to \infty$, each cell moves in a fixed direction for the duration of the simulation. 

Mechanically stable packings of deformable particles (with bending energy in Eq.~\ref{eq:shape_energy}) in the zero-activity limit possess mean shape parameters that are less than $\calAzero$ since the bending energy and interparticle contacts cause rounding of the deformable particles. In Fig.~\ref{fig:DPM} (c), we show the mean shape parameter of deformable particle packings for Model $1$, $\calAzero'$, as a function of $\calAzero$ and $\epsilon_b/(k_l\sigma^2)$. In general, the differences between $\calAzero'$ and $\calAzero$ are small with $(\calAzero-\calAzero')/\calAzero<10\%$ for the values of $\epsilon_b/(k_l\sigma^2)$ used in these studies.

\subsection{Model 2: Deformable Particle Model with Adaptive Preferred Perimeter}

\label{dpm_methods_plastic}

The shape-energy function $U_{\mu,\text{shape}}$ for cell $\mu$ and total energy $U$ for a collection of $N$ cells using Model $2$ will have the same forms as those in Model $1$. However, for the second model of motile epithelial cells in confluent monolayers, we will allow the preferred perimeter $p_{\mu0}$ of each cell to vary based on the local forces it experiences. We first generate a mechanically stable packing of cells using the same protocol as that described in Sec.~\ref{sec:dpm_methods_fixed}. We start with non-overlapping cells at a fixed $\calAzero$ and $\phi=0.4$ and increase $\phi$ in small steps followed by energy minimization until reaching $\phi=0.95$.

As for Model $1$, we induce cell motion by adding an external force $f_0/N_v$ to each vertex, as well as a drag force so that the monolayer can reach a steady state.  The key difference between Models $1$ and $2$ is that for Model $2$ we allow the preferred perimeter to adapt to local forces:
\begin{equation}
    \frac{dp_{\mu0}}{dt} = \frac{1}{\tau_p }(p_{\mu }-p_{\mu0}),
    \label{eq:plasticity_equation}
\end{equation}
 where $\tau_p=\eta/
 k_l$ is a timescale that controls the rate of relaxation of $p_{\mu0}$, and $\eta$ is proportional to a damping coefficient with units of $\sqrt{m k_l}$. The cell can thus release its stress via the relaxation of $p_{\mu0}$. As $\tau_p \to \infty$, we recover Model 1 with fixed $p_{\mu0}$, while $\tau_p \to 0$ allows the preferred perimeter to change instantaneously to the current perimeter. 
 
\begin{figure}
    \centering
    \includegraphics[width=\columnwidth]{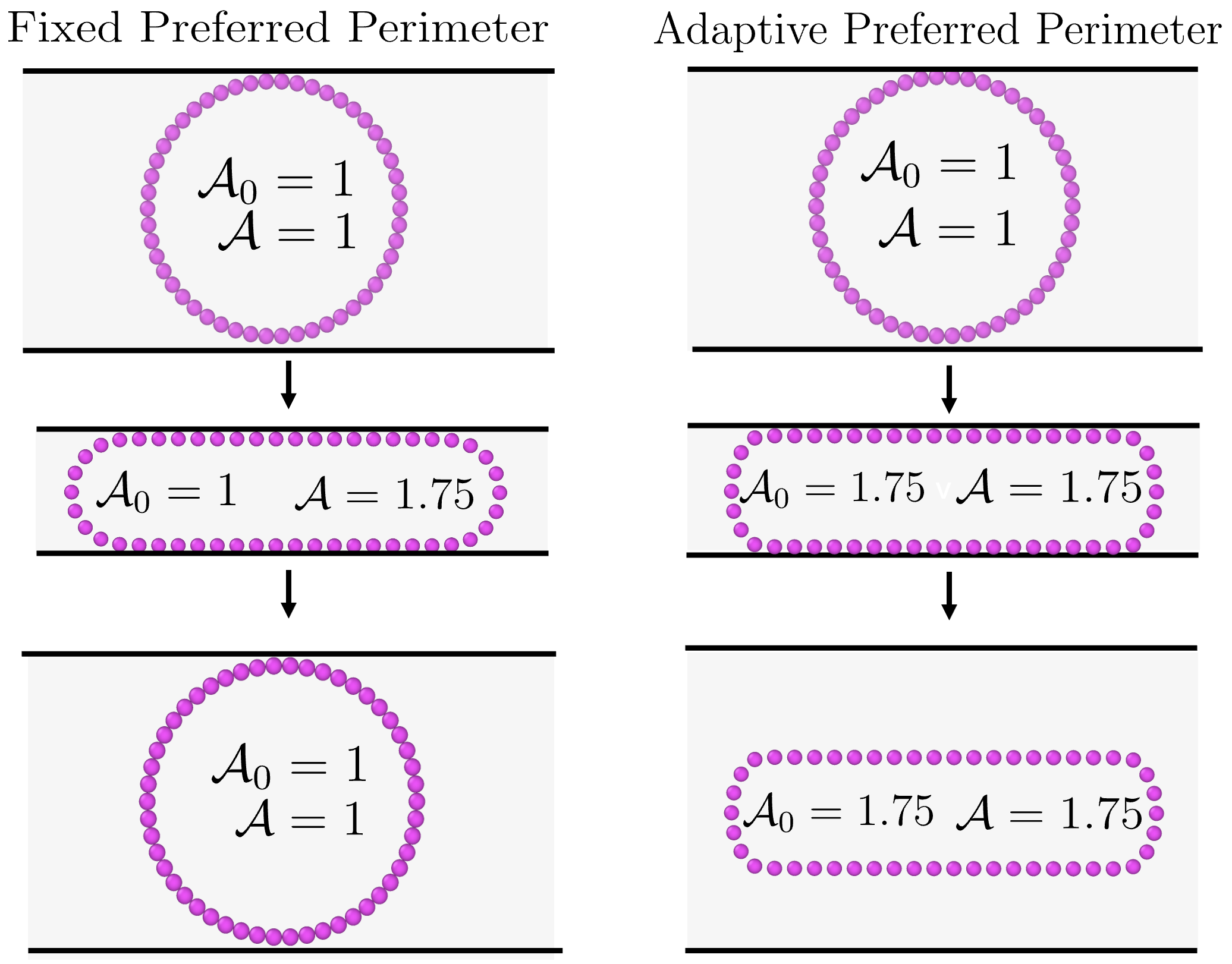}
    \caption{(left column; Model $1$) A deformable particle with fixed preferred perimeter (undeformed diameter $D_0$ and shape parameter ${\cal A}_0 = 1.0$) is compressed between two rigid parallel plates so that they have separation $h/D_0=0.47$. The particle deforms from $\calA=1$ to $1.75$, and returns to $\calA=1$ when the applied compression is removed ($h/D_0=1.0$). (right column; Model $2$) A deformable particle with an adaptive preferred perimeter (initial ${\cal A}_0=1.0$) is compressed between two rigid parallel plates to ${\cal A}=1.75$ at $h/D_0 = 0.49$. As it is compressed, the perimeter relaxes to a new value, causing the preferred shape parameter to change from ${\cal A}_0=1.0$ to $1.75$. After the applied compression is removed, the deformable particle remains in the shape that it had during the compression.}
    \label{fig:plastic_model}
\end{figure}

In contrast to Model 1 for which there is a harmonic restoring force for maintaining a given $\calAzero$, Model 2 allows cells to deform to a new preferred perimeter (and $\calAzero$) when they are strained.  In Model $2$, cells lose memory of their initial $\calAzero$ as they are deformed, and their average shape parameter becomes an emergent property of the cells' collective motion.  In Fig. \ref{fig:plastic_model}, we illustrate these points through simulations of single deformable particles undergoing uniaxial compression for Models $1$ and $2$. For Model $1$, we compress a deformable particle with $\calAzero=1$ until it has ${\cal A} = 1.75$, and the ratio of the final distance between the compression plates to the initial diameter of the deformable particle is $h/D_0=0.47$. We then remove the compression (i.e. set $h/D_0=1.0$) and the particle returns to its original shape. For Model $2$, we again compress the deformable particle to $\calAzero=1.75$, but now $h/D_0=0.49$ since the particle can deform more easily at a given $h$. After the compression is removed (with $h/D_0 =1.0$), the deformable particle maintains the compressed shape with ${\cal A}= 1.75$ and a new value for ${\cal A}_0 = 1.75$.

\subsection{Selection of the Simulation Parameters}
\label{sec:simulation_parameters}

\begin{figure*}[htbp]
    \centering
    \includegraphics[width=\textwidth]{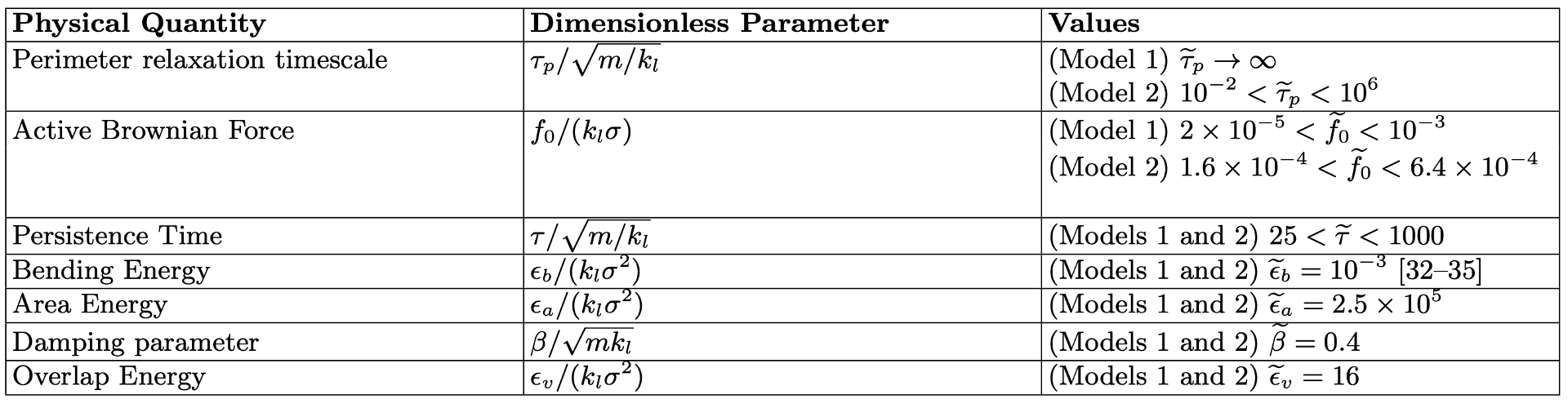}   
    
    \captionof{table}{Listing of the (left) physical quantities, (center) associated dimensionless parameters, and (right) the values that were used in the deformable particle simulations of epithelial cell monolayers for Models $1$ and $2$.}
    \label{table:simulation_parameters}
\end{figure*}

In this subsection, we describe the range of simulation parameters that we investigate for the simulations of epithelial cell monolayers using Models $1$ and $2$. These models include seven dimensionless parameters; we constrain four of them to values from the experiments and vary the remaining three to determine their effects on the cell shape parameter distribution. For Models $1$ and $2$, we sweep over the dimensionless active Brownian force ${\widetilde f}_0 = f_0
/(k_l\sigma)$ and persistence time ${\widetilde \tau}=\tau/\sqrt{m/k_l}$, which vary the degree of activity of the cells. In Model $2$, we also sweep over the dimensionless perimeter relaxation timescale $\tau_p/\sqrt{m/k_l}$. The ranges of the dimensionless parameters are provided in Table \ref{table:simulation_parameters}.

We fixed the dimensionless area energy ${\widetilde \epsilon}_a = \epsilon_a/(k_l\sigma^2)=2.5\times10^5$ for both models so that the area fluctuations satisfy $\Delta a_{\mu}/a_{\mu 0} < 10^{-4}$.  We set the dimensionless overlap energy $\widetilde\epsilon_v=\epsilon_v/(k_l\sigma^2)=16$ so that the overlap between vertices is less than $10^{-3} \sigma$. Previous experimental studies of isolated membranes and mammalian cells have reported that the membrane bending modulus and cortical tension occur in the range $10^{-21} < \epsilon_b < 10^{-19}$ J \cite{nagle2013introductory} and  $10^{-5} < k_l < 10^{-2}$ N/m \cite{peukes2014direct, moazzeni2021single, dai1999myosin}. The diameter of a vertex $\sigma$ in the simulations corresponds to approximately one tenth of a cell diameter, or $\sim 1\mu$m. Thus, we set $\epsilon_b/(k_l\sigma^2)=10^{-3}$ for the simulations, which falls within the experimental range $10^{-7} < {\widetilde \epsilon}_b < 10^{-2}$.

\section{Results}
\label{results_section}

In this section, we begin by describing the shape parameter distributions $P(\calA)$ obtained from the cell segmentation of datasets (1) and (3), i.e. the low-density islands of MDCK and HaCaT epithelial cells.  To calculate $P(\calA)$, we bin the shape parameters of all cells over all frames in a dataset to generate a spatially and temporally averaged distribution.  We find that $P(\calA)$ is broad and positively skewed, in contrast to the narrow distribution of shape parameters found for jammed, static packings of epithelial cells. We focus on low-density cell islands in which the cells are moving, i.e. the cell displacements in the low-density islands are typically greater than a cell diameter over the timescale of the experimental measurements, in contrast to those for the high-density islands~\cite{saraswathibhatla2020tractions, angelini2011glass}.  We show that $P({\cal A},t)$, the shape parameter distribution at time $t$ for the low-density islands of MDCK epithelial cells, reaches a steady-state after a few minutes into the experimental measurements (Appendix B). We then compare the spatially and temporally averaged $P({\cal A})$ to the shape parameter distribution obtained for a single cell $\mu$ averaged over time.  We seek to identify the key biophysical mechanism that gives rise to the broad, positively skewed shape parameter distribution by studying two models for motile epithelial cell monolayers: (1) the deformable particle model (DPM) for cells with fixed preferred shape parameters and (2) the DPM model for cells with an adaptive preferred perimeter. We determine the possible parameter regimes for both models for which the predicted $P({\cal A})$ recapitulates the shape parameter distribution for the low-density islands of motile epithelial cell monolayers. 

\begin{figure}[tb]        

    \includegraphics[width=\columnwidth]{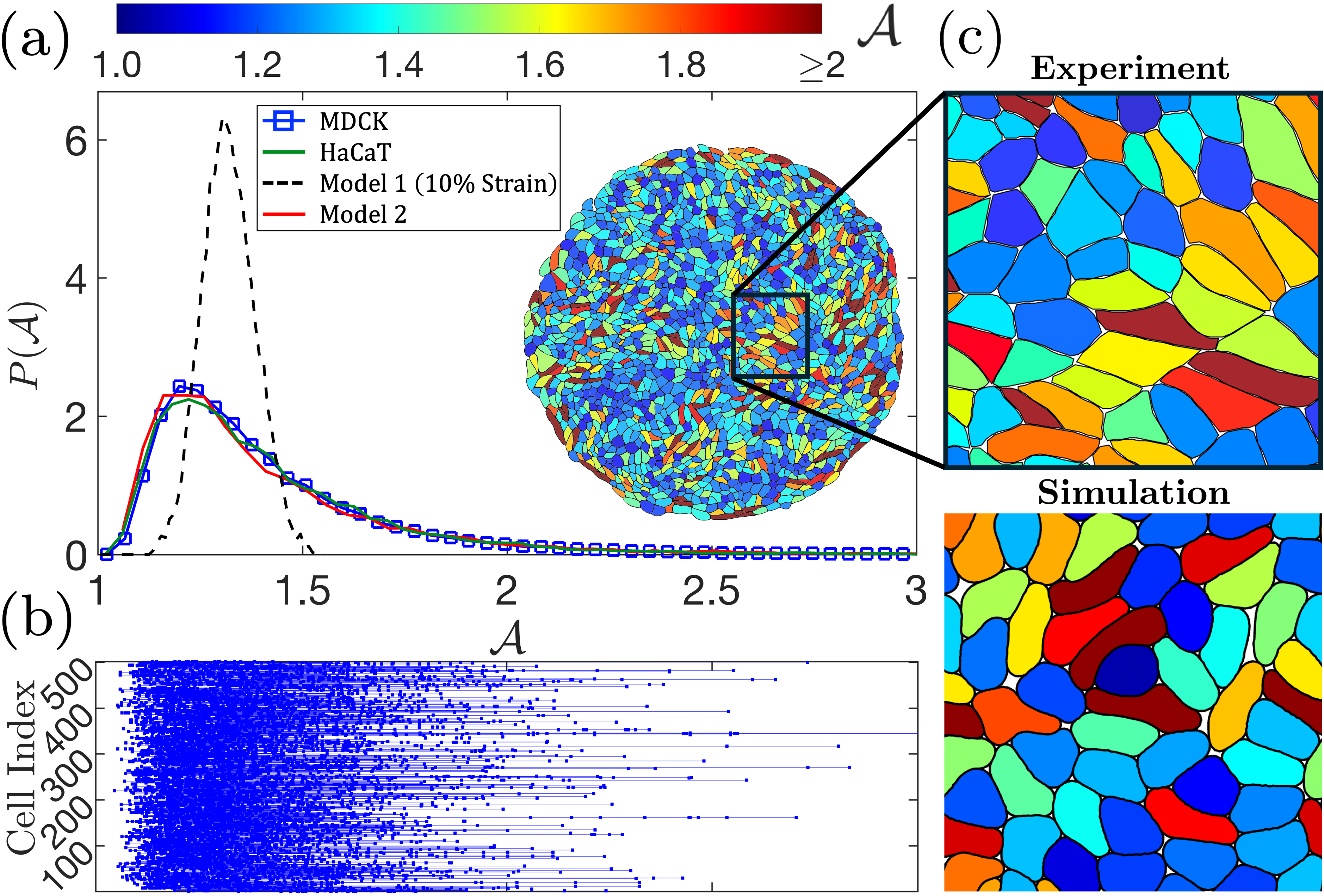}
    
    \caption{(a) The distribution of shape parameters $P(\mathcal{A})$ from low-density islands of MDCK (blue squares) and HaCaT (green solid line) cells and deformable particle model simulations with fixed preferred shape parameter ${\cal A}_0 =1.4$ and at most $10 \%$ perimeter strain (Model 1, black dashed line) and adaptive preferred perimeter (Model 2, red solid line). Inset: Full segmented MDCK cell monolayer with cells colored by $\calA$ increasing from blue to red. (b) Each horizontal line and associated points represent the range of shape parameters for individual MDCK cells (labeled from $1$ to $500$) that were sampled over the $180$-minute experiments using the same horizontal scale as (a). (c) (top) Region of a segmented MDCK cell monolayer and (bottom) snapshot from the deformable particle simulations of Model 2 in (a) with cells colored by the shape parameter.}
    \label{fig:MDCK}
\end{figure}

In Fig.~\ref{fig:MDCK} (a), we show  $P(\calA)$ (averaged over cells and time) for the low-density islands of MDCK and HaCaT epithelial cells.  We find that the distributions for MDCK and HaCaT cells are similar with a normalized root-mean-square error $< 5\%$. $P({\cal A})$ for the MDCK and HaCaT epithelial cells peaks at $\calA \approx 1.2$, which is above the minimum ${\cal A} \approx 1.15$ required for confluent cell monolayers~\cite{boromand2018jamming}, and is positively skewed with a long tail that extends to $\calA > 2$. As shown in Appendix C, $P(\calA)$ is well-fit by a gamma distribution, whose origin has been shifted to account for the fact that $\calA_{\text{min}}=1$. We characterize the form of $P({\cal A})$ using its first three moments: the mean $\meancalA \approx 1.41$, normalized variance $\varcalA=\sigma_{\cal A}^2/\langle \mathcal{A}\rangle^2 \approx 0.043$, and skewness $\skewcalA=\langle (\calA-\langle\calA\rangle)^3\rangle/\sigma_{\cal A}^3  \approx 2.45$, where $\sigma_{\cal A}^2 = \langle (\mathcal{A}-\langle \mathcal{A}\rangle)^2\rangle$.

We first describe the predictions for $P({\cal A})$ from deformable particle simulations of Model $1$. In one extreme, we can assume a heterogeneous collection of cells each with distinct preferred shape parameters ${\cal A}_{\mu0}$ and large values for the perimeter stiffness $k_l$. If the populations of cells with $A_{\mu0}$ match $P({\cal A})$, we can recover the shape parameter distribution from the low-density islands of MDCK and HaCaT epithelial cells. To test this hypothesis, we analyzed the shape parameters for $500$ individual cells in low-density islands of MDCK cells as a function of time. We find that each cell samples a wide range of shape parameters with a variance that is comparable to that of the full distribution $P(\calA)$. (See Fig. \ref{fig:MDCK} (b).) Thus, epithelial cells do not strongly fix their shape parameter to a preferred value. 

    \begin{figure}[h!]
		\includegraphics[width=\columnwidth]{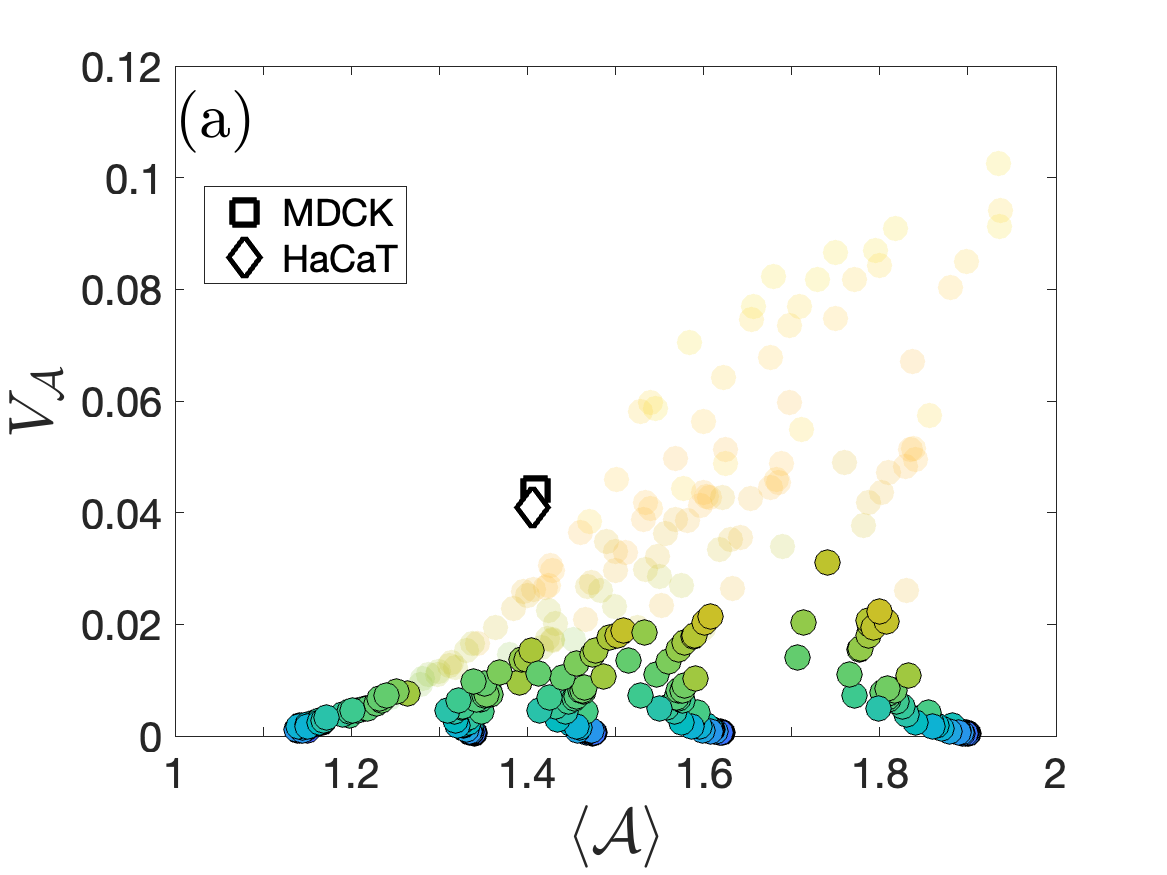}
        \includegraphics[width=\columnwidth]{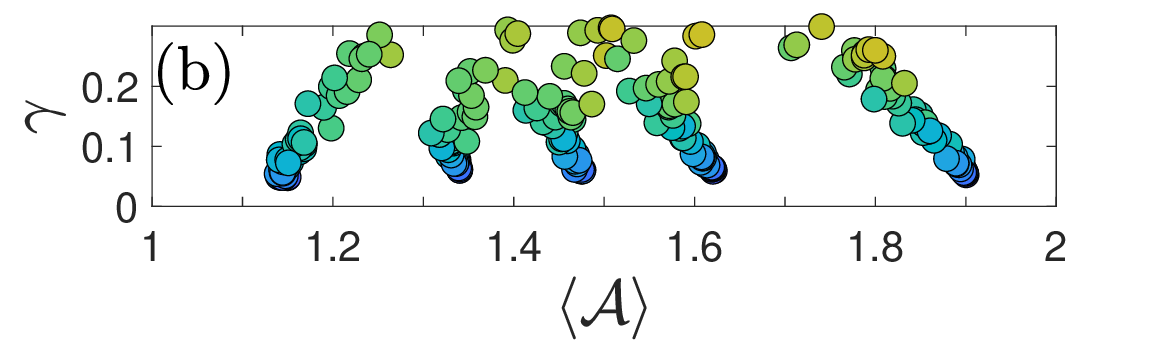}
        \includegraphics[width=\columnwidth]{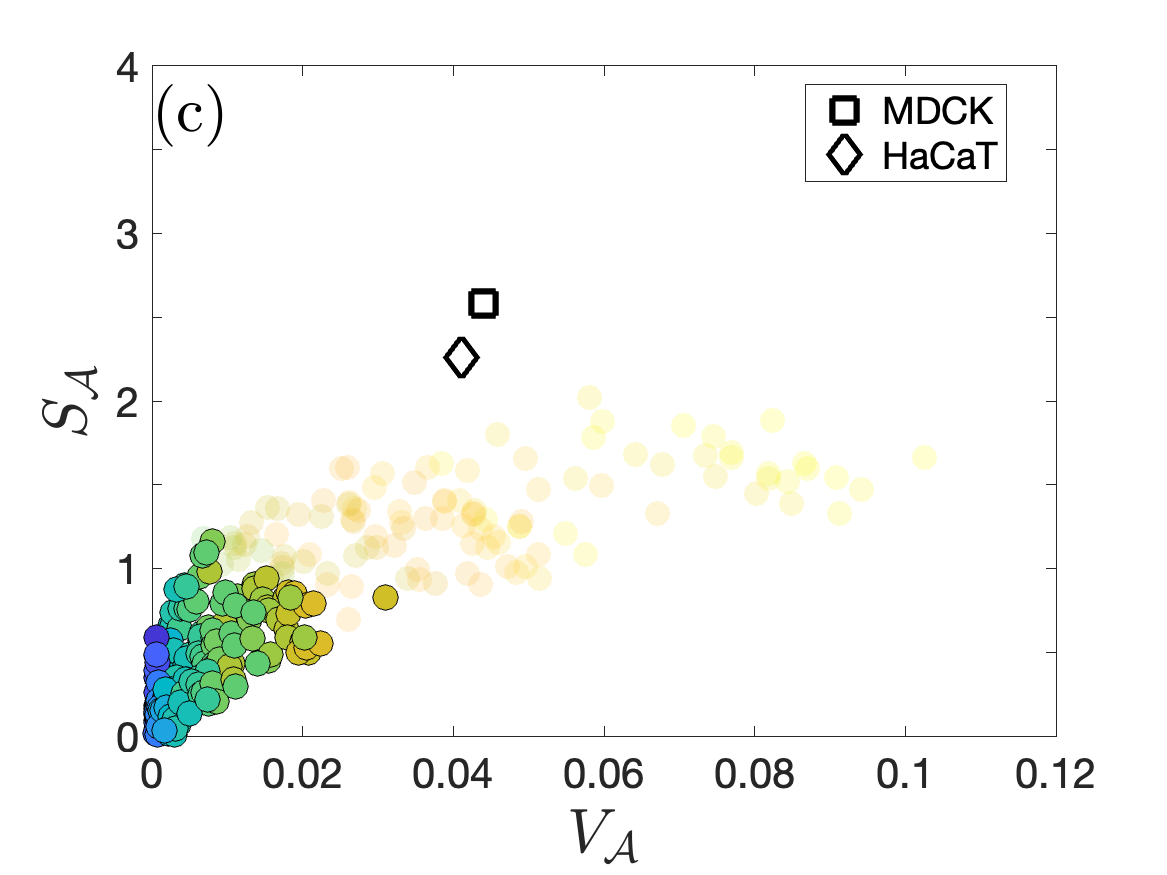}
        \includegraphics[width=\columnwidth] {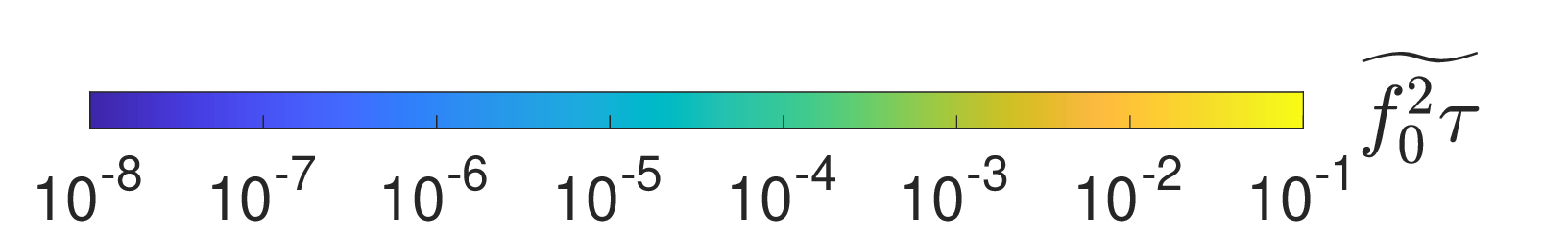}
		\caption{(a) The mean $\meancalA$ and normalized variance $\varcalA$ of the shape parameter distribution from deformable particle simulations of Model 1 with fixed $\calAzero'=1.14$, $1.33$, $1.46$, $1.6$ and $1.88$. The color of the data points indicates the degree of cell activity $\widetilde{f_0^2 \tau}$ increasing from blue to yellow. The faded circles represent simulations whose maximum perimeter strain $\gamma = |p-p_0|/{p_0} > 0.3$. (b) $\gamma$ versus $\meancalA$ from simulations with fixed $\calAzero$ for activity values that yield $\gamma <0.3$. (c) The skewness $\skewcalA$ plotted versus $\sigma^2_{\cal A}$ for the data in (a). $\meancalA$ and $\varcalA$ for the low-density islands of MDCK and HaCaT cells are also included (white square and diamond). }
		\label{fig:static_cal_A}
	\end{figure}

In another extreme, we can consider deformable particle simulations of Model $1$ using a single value for the preferred shape parameter ${\cal A}_{\mu 0} = {\cal A}_0$, such that ${\cal A}$ for each cell fluctuates about ${\cal A}_0$ from deformations in the perimeter that occur during cell motion.  We carry out simulations of Model 1 as a function of activity $\widetilde{f_0^2\tau} =f_0^2 \tau /( \sigma^2\sqrt{k_l^3} )$ and five values of $\calAzero=1.15$, $1.4$, $1.55$, $1.7$, and $2$.  The mean shape parameter of the deformable particle packings in the limit of zero activity is $\calAzero'$  = $f(\calAzero, \epsilon_b/(k_l\sigma^2))$,  where $f$ is a decreasing function of $\epsilon_b/(k_l\sigma^2)$ as shown in Fig. \ref{fig:DPM} (c). For the simulations that we carried out for Model $1$, $\calAzero'=1.14$, $1.33$, $1.46$, $1.6$ and $1.88$, which are all within 10\% of $\calAzero$. Using $\calAzero'$ as a reference allows us to isolate the effect of cell motion on the shape parameter distribution.  We use the combination $\widetilde{f_0^2\tau}$ to quantify the cell activity since it is proportional to the effective temperature in the small-$\tau$ limit \cite{marcetti2012,fily2014freezing} and $\varcalA$ and $S_{\cal A}$ vary monotonically with $\widetilde{f_0^2\tau}$. 

In Fig. \ref{fig:static_cal_A} (a), we show that $\langle \calA \rangle \sim {\cal A}_0'$ for small values of the activity, $\widetilde{f_0^2\tau} \lesssim 10^{-4}$, but $\varcalA$ and $\meancalA$ increase for large values of activity. Increases in activity yield increases in the perimeter strain $\gamma=|p-p_0|/p_0$ as shown in Fig. \ref{fig:static_cal_A} (b).  Note that at small strains $\meancalA$ increases with activity for $\mathcal{A}_0'=1.14$, which emphasizes that cells with small shape parameters must increase their perimeters to move. However, $\meancalA$ {\it decreases} with activity at small strains for $\calAzero' \gtrsim 1.3$, which implies that cells with large shape parameters can interlock during motion, and thus they must decrease their perimeters to move efficiently. 

To identify the biophysically realistic parameter range for cell activity using Model $1$, we fade the colors of the data points in Fig. \ref{fig:static_cal_A} that correspond to perimeter strains greater than $30\%$, which is significantly above the cells' elastic strain limit \cite{fernandez2008single,cordes2020prestress, chugh2018actin}.  When $\calAzero'=1.14$ and $\widetilde{f_0^2\tau} \approx 10^{-2}$, $\meancalA$ and $\varcalA$ for Model $1$ are closest (i.e. within 5\%) to the experimental values for the MDCK and HaCaT cells. However, to achieve these values for $\meancalA$ and $\varcalA$, the cells are strained well beyond their elastic limit.  If we choose values for $\calAzero$ and $\widetilde{f_0^2\tau}$ that yield realistic strains $\gamma \lesssim 10\%$, $P(\calA)$ is much narrower than the distributions observed for the MDCK and HaCaT cells, as shown in Fig.~\ref{fig:MDCK} (a).

We also calculate the skewness $\skewcalA$ of the shape parameter distributions as a function of $\varcalA$ for Model $1$ in Fig. \ref{fig:static_cal_A} (c).  We find that the skewness plateaus at $\skewcalA <2$ in the large activity limit, which is well below the skewness of $P({\cal A})$ for MDCK and HaCaT cells. Moreover, when $\gamma < 0.3$, the skewness of $P({\cal A})$ for Model $1$ is even smaller with $\skewcalA \lesssim 1.2$. In addition, the properties of $P({\cal A})$ obtained from Model $1$ depend strongly on the reference shape parameter $\calAzero'$, while $P({\cal A})$ for the epithelial cell monolayers are robust for different cell types and replicates. (See Appendix B.)   Thus, Model $1$ with a fixed preferred shape parameter does not explain  $P(\calA)$ observed for epithelial cell monolayers. 

    \begin{figure*}[tb]
    \centering
    \begin{minipage}{0.49\textwidth}
        \includegraphics[width=\linewidth]{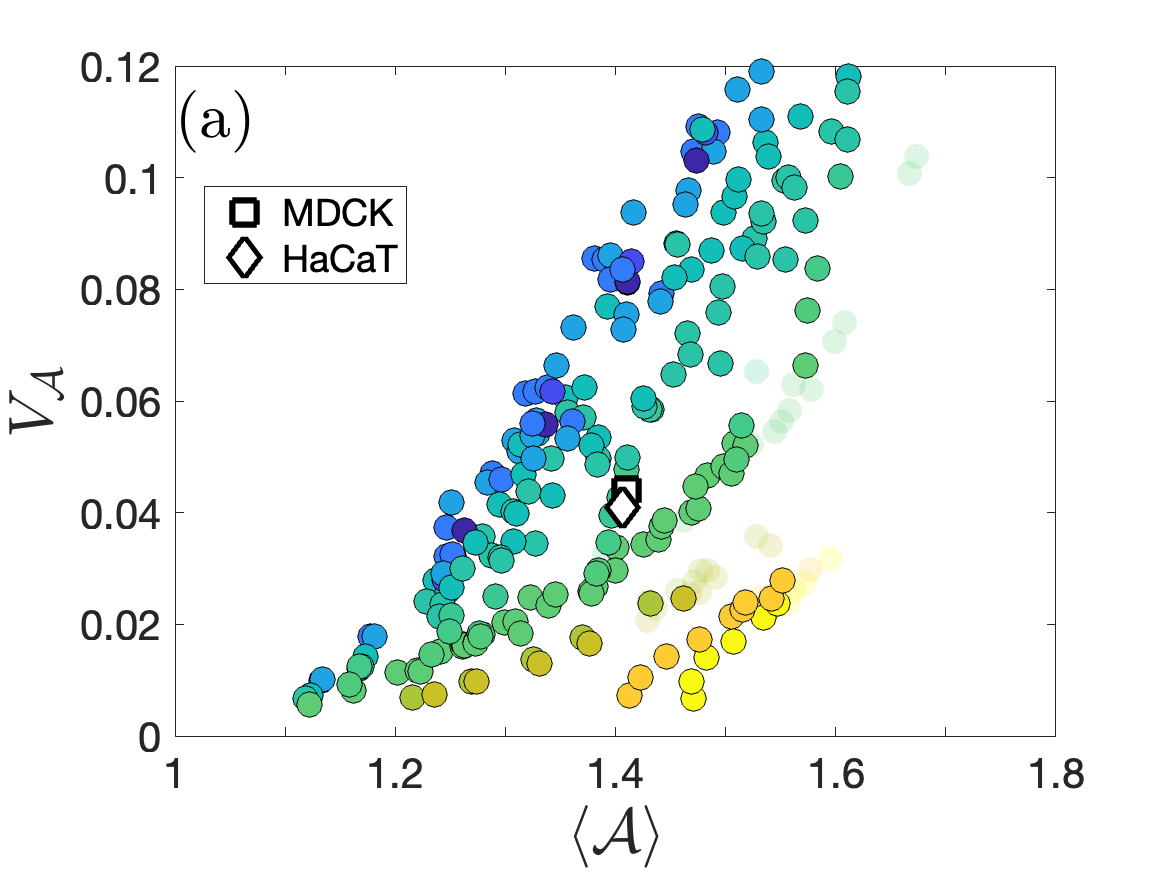}
        \label{fig:plastic_mu1_mu2}
    \end{minipage}
    \hfill
    \begin{minipage}{0.49\textwidth}
        \includegraphics[width=\linewidth]{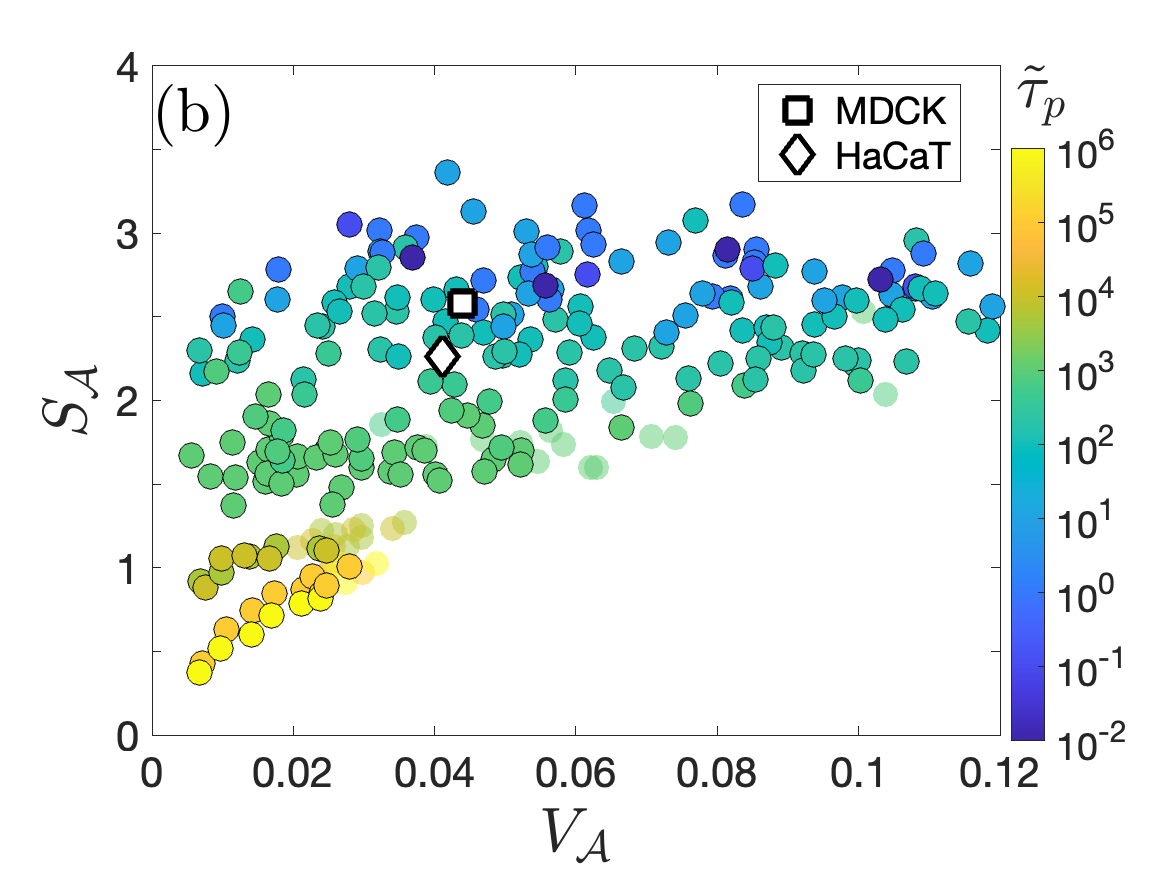}
        \label{fig:plastic_mu2_mu3}
    \end{minipage}
    
    
    \caption{(a) The mean shape parameter $\meancalA$ and normalized variance $\varcalA$ for deformable particle simulations with an adaptive preferred perimeter (Model $2$). The color of each point represents the perimeter relaxation timescale $\widetilde\tau_p$ with increasing values from blue to yellow. (b) The skewness $\skewcalA$  plotted versus $\varcalA$ for the same data in (a). The squares and diamonds represent the corresponding shape parameter moments for the MDCK and HaCaT cells, respectively.}
    \label{fig:plastic_cal_A}
\end{figure*}

We now describe the results from deformable particle simulations of Model $2$, where the preferred perimeter of each cell $p_{\mu0}$ can evolve in response to local forces. In Fig. \ref{fig:plastic_cal_A} (a), we show $\varcalA$ as a function of $\meancalA$ for Model 2 while varying $\widetilde{f_0^2\tau}$ and the perimeter relaxation timescale $\widetilde{\tau}_p$. We find that there are several points (near $\widetilde\tau_p\approx 10^{3}$) where $\meancalA$ and $\varcalA$ are similar to those for the epithelial cell monolayers.  We select the parameters for the closest data points to those for the MDCK and HaCaT cells in Fig.~\ref{fig:plastic_cal_A} (a) and plot the corresponding shape parameter distribution in Fig. \ref{fig:MDCK} (a). $P({\cal A})$ from the Model $2$ simulations and experiments have a normalized root-mean-squared error of $\lesssim 5\%$. This small error is reflected in Fig. \ref{fig:MDCK} (c), where we compare snapshots of the Model $2$ simulations and a small region within a low-density MDCK island. 

In Fig. \ref{fig:plastic_cal_A} (b),  we show $\skewcalA$ as a function of $\varcalA$ for the Model $2$ simulations, and again $\skewcalA$ is within $10\%$ of the values for epithelial cells when $\widetilde\tau_p\approx10^3$. Having an adjustable $p_{\mu0}$ allows cells to deform to larger shape parameters and maintain their shapes, resulting in larger values of $\skewcalA$ that match those of epithelial cells. In addition, Model 2 captures the epithelial shape parameter distributions at strains $\gamma$ that are within the elastic limit (indicated by the lack of faded points in Figs. \ref{fig:plastic_cal_A} (a) and (b) with $\gamma>30\%$).

\begin{figure}[b]
    \centering
    \includegraphics[width=\columnwidth]{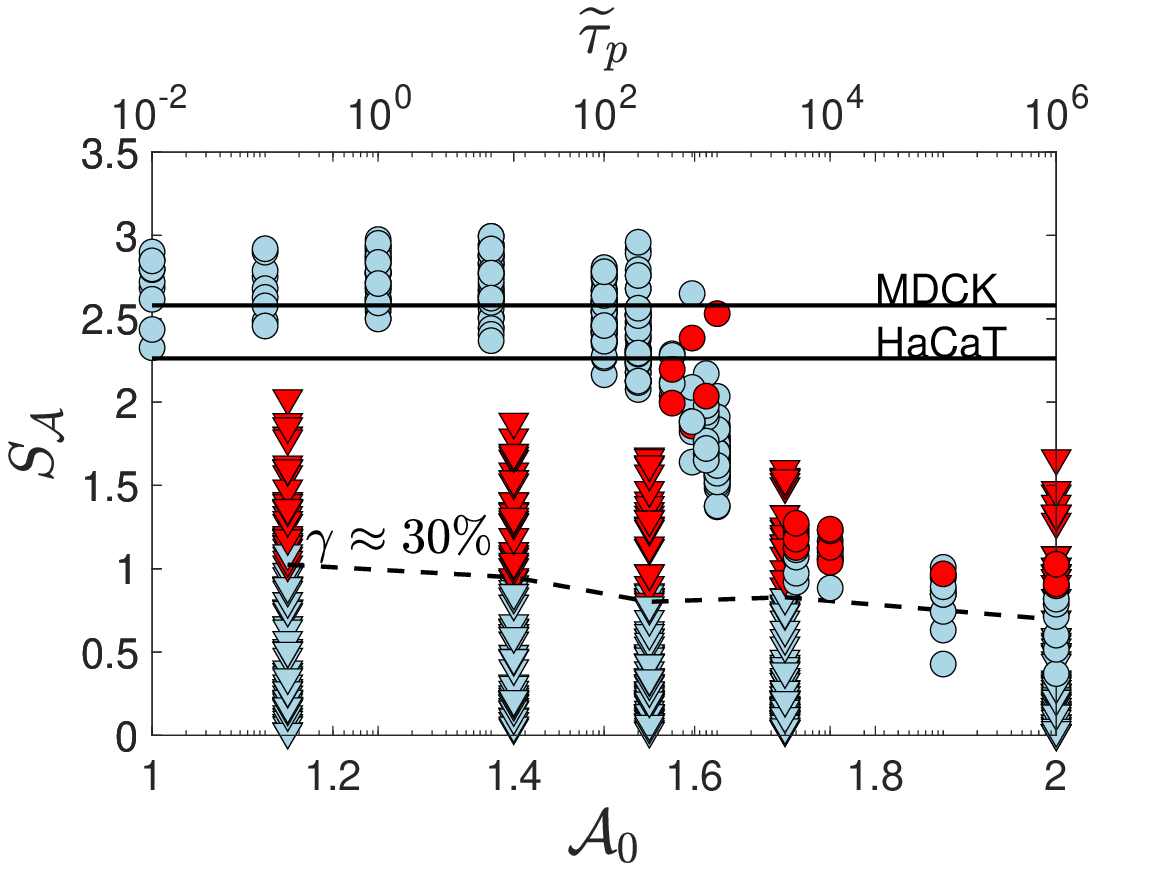}
    \caption{$\skewcalA$ plotted as a function of $\calAzero$ for Model 1 (triangles, bottom axis) and as a function of $\widetilde{\tau}_p$ for Model 2 (circles, top axis). The red points indicate the simulations for which the perimeter strain $\gamma>30\%$. The labeled black horizontal lines indicate $\skewcalA$ for MDCK and HaCaT cells and the dashed black line indicates the skewness boundary above which Model 1 simulations have $\gamma > 30\%$.}
    \label{fig:robustness}
\end{figure}

All of the shape parameter distributions for Model 2 were generated from a single, initial configuration of cells with $\calAzero'=1.6$. Thus, for Model $2$, we find that there is a well-defined steady state $P(\calA)$ for a given ${\widetilde \tau}_p$ and $\widetilde{f_0^2\tau}$ that does not depend on the input parameter $\calAzero'$. We showed in Fig.~\ref{fig:plastic_cal_A} that the Model $2$ simulations and cell monolayers have similar values for $\meancalA$, $\varcalA$, and $\skewcalA$ for ${\widetilde \tau}_p = 10^{3}$.  Are the moments of the shape parameter distribution for the Model $2$ simulations sensitive to ${\widetilde \tau}_p$?  In Fig. \ref{fig:robustness}, we show $\skewcalA$ as a function of $\calAzero$ for Model 2 (circles) and find that when $\widetilde\tau_p\lesssim10^2$, $\skewcalA$ remains within 20\% of the MDCK and HaCaT cell values. In contrast, when $\widetilde\tau_p \gg 10^2$, $\skewcalA$ decreases to $\sim 1.0$, which represents the fixed-$\calAzero$ limit (Model $1$; triangles). 

\section{Discussion}
\label{discussion_section}

In low-density confluent monolayers, epithelial cells are highly motile and dynamically adjust their shapes based on the mechanical forces they experience in their local environments. To quantify the shape fluctuations of epithelial cells, we calculate the shape parameter $\calA$ of each cell in MDCK and HaCaT epithelial cell monolayers as a function of time. Averaging $\calA$ of each cell over time and different monolayers, we find that the cell shape parameter distribution $P(\calA)$ is broad and positively-skewed with mean $\langle \calA \rangle \sim 1.4$. This shape parameter distribution is robust, since the average over cells at a given time is similar to the average of a single cell over time and the distributions are the same for different cell lines. 

To understand the underlying biophysical mechanisms that give rise to the shape parameter distribution in confluent epithelial cell monolayers, we analyze two computational models of dense packings of active deformable particles. We find that deformable particle simulations with fixed preferred shape parameter (Model $1$) cannot recapitulate the shape parameter distribution found in low-density MDCK and HaCaT epithelial cell islands without large perimeter strains beyond the elastic limit. In addition, the shape parameter distributions from Model $1$ are highly sensitive to the input preferred shape parameter, whereas confluent epithelial cell islands reach the same shape parameter distributions for different cell lines and experimental set-ups (Appendix B, Fig.~\ref{fig:single_islands}). However, when we model epithelial cells as deformable particles with an adaptive preferred perimeter (Model $2$), we can recover the shape parameter distribution of the low-density MDCK and HaCaT epithelial cell islands, and the properties of the distribution are insensitive to the precise value of the perimeter relaxation time scale ${\widetilde \tau}_p$ as long as it is sufficiently small. The success of Model $2$ emphasizes that the shape parameter distribution of confluent, epithelial cell monolayers is an emergent physical property of motile, confluent cell monolayers.

The conclusions of this study contrast with the previous paradigm that cells choose a specific value for the preferred shape parameter $\calAzero$. In the previous paradigm, $\calAzero$ controls the cell monolayer pre-stress, and hence the transition from solid- to fluid-like behavior~\cite{bi2015density,bi2016voronoi}. Our work emphasizes that the cell shape parameter distribution emerges from the cell motion in fluidized confluent monolayers. Thus, what variables determine whether confluent cell monolayers are solid- or fluid-like? Another important variable in confluent cell monolayers is the cell number density \cite{angelini2011glass}. Our image segmentation shows that high-density epithelial cell islands (both MDCK and HBEC) have much narrower shape parameter distributions compared to low-density islands. (See Appendix D.) Previous work also shows that these high-density cell islands are more solid-like with fewer cell rearrangements and smaller cell displacements than the fluidized low-density cell islands \cite{saraswathibhatla2020tractions}. In addition, the result that low-density MDCK and HaCaT epithelial cell islands exhibit similar shape parameter distributions indicates that the cell number density may serve as a control variable that distinguishes the shape distributions for fluid- versus solid-like confluent cell monolayers. Prior work has also shown that cell traction, and hence cell motility, decreases with density, which may explain why cells become dynamically arrested \cite{saraswathibhatla2020spatiotemporal}.  In future work, we will carry out deformable particle simulations of Model $2$ with coupling between the cell density and the mechanical and motility properties of the cells to evaluate the shape parameter distribution across the solid- to fluid-like transition. 

There are several possible biological mechanisms for the evolution of the preferred perimeter (or area) in Model $2$. One mechanism for perimeter relaxation in epithelial cells is the force-dependent folding and unfolding of membrane reservoirs (or caveolae) \cite{gervasio2011caveolae, fernandez2008single}. In addition, variation in membrane tension can modulate the rates of endocytosis and exocytosis, allowing vesicles to merge with and be removed from the membrane \cite{apodaca2002modulation, sheetz2001cell}. Increases in cell perimeter are also accompanied by remodeling of the cell cortex, and thus the timescale for this remodeling may control the perimeter relaxation process \cite{kelkar2020mechanics}.  Our study focuses on perimeter relaxation; however, if cells modulate their preferred area via a mechanism similar to Eq. \ref{eq:plasticity_equation}, we would obtain the same results. Epithelial cells can also exchange fluid via gap junctions \cite{zehnder2015cell}, such that the area of individual cells can change but the total area of the cell monolayer remains constant. In future work, we will calculate $P({\cal A})$ in deformable particle systems that exchange cell area, but the total cell area is held fixed. We conclude that cell shape parameter adaptation represents a general biophysical mechanism, which gives rise to the broad, positively-skewed shape parameter distribution observed in fluidized confluent epithelial cell monolayers. 

\section*{Acknowledgments}

We acknowledge support from NSF grant No. POLS-2102789 (G.G. and C.S.O.) and NIH Grant Nos. T32GM145452 (E.Y.M. and C.S.O) and R35GM151171 (J.J.S. and J.N.). This work was also supported by the High Performance Computing facilities operated by Yale’s Center for Research Computing. We also thank Dong Wang for helpful discussions and Grace Yang and Matilda Ryder for their assistance in cell segmentation of the HBEC monolayers.  

\section*{Appendix A: Experimental Conditions for Epithelial Cell Monolayers}

Madin-Darby Canine Kidney (MDCK) type II cells and HaCaT cells (dataset $1$-$3$ in Table~\ref{table:image_details}) were maintained in Dulbecco's Modified Eagle's Medium (DMEM) with $10\%$ Fetal Bovine Serum (FBS), and $1\%$ penicillin-streptomycin at $37^\circ$C with $5\%$ CO$_2$ in an incubator. During the time-lapse imaging experiments, both the HaCaT and MDCK cell islands were seeded in low-glucose DMEM containing $1\%$ FBS and $1\%$ penicillin-streptomycin. Cells were seeded onto polyacrylamide substrates with a Young's modulus of $6$ kPa and a thickness of $120$ {\textmu}m  using published concentrations of acrylamide and bis-acrylamide \cite{tse2010preparation}. Using micro-patterning \cite{saraswathibhatla2020spatiotemporal, saraswathibhatla2020tractions}, cells were seeded onto $1$ mm diameter islands of collagen I, which enables control of the cell density. Phase contrast, time-lapse images of the prepared HaCaT and MDCK cell islands were taken using the phase contrast mode of an Eclipse Ti microscope (Nikon) using a 10X magnification objective with a numerical aperture of $0.5$ (Nikon) and an Orca Flash 4.0 camera (Hamamatsu). Images of the HaCaT and MDCK cell islands were collected every $10$ min and the imaging environment was maintained at $37^\circ$C and $5\%$ CO$_2$ for all experiments.

Madin–Darby Canine Kidney (MDCK) type II epithelial cells (dataset ($4$) in Table~\ref{table:image_details}) stably expressing ZO-1–GFP \cite{lee2017polymer} were maintained in DMEM supplemented with $10\%$ FBS and $1\%$ penicillin–streptomycin at $37^\circ$C and $5\%$ CO$_2$. An ibidi Culture-Insert ($\textmu$-Dish $35$ mm, Cat. 81176) was placed in a $35$ mm dish, and seeded with $56,000$ cells in $70$ $\textmu$L. To prevent dehydration, $2$ mL of additional medium was added around the insert. After $20$ hours, the insert was removed to initiate collective migration, generating a dense, actively migrating MDCK monolayer. Imaging was performed on a ZEISS Axio Vert.A1 microscope using a 10X objective.

Primary human bronchial epithelial cells (HBECs, datasets (5) and (6) in Table~\ref{table:image_details}) from asthmatic and non-asthmatic donors were cultured under air–liquid interface conditions \cite{atia2018geometric}. At days $6, 10, 14$, and $20$, monolayers were fixed with $4\%$ paraformaldehyde and stained with Alexa-488 conjugated with phalloidin to visualize F-actin. Wide-field fluorescence images of the apical plane were acquired on a Leica DMI8 microscope using either a 40X or 63X oil-immersion objective.

\section*{Appendix B: Averaging of Experimental Shape Parameter Distributions}
\label{sec:mdck_datasets}

In this appendix, we discuss the calculations of the shape parameter distributions for epithelial cell monolayers (MDCK and HaCaT) for which we have time-lapse images. For the low-density islands of MDCK epithelial cells, we have four time-lapse imaging datasets, each containing $18$ images of the islands taken $10$ minutes apart. The shape parameter of each cell within a single island at a given time $t$ is calculated and averaged over cells to generate $P(\calA, t)$. We can also average $P(\calA,t)$ across all time frames and islands to generate $P(\calA)$, as shown in Fig.~\ref{fig:MDCK} (a).

\begin{figure}[tb]
    \includegraphics[width=\columnwidth]{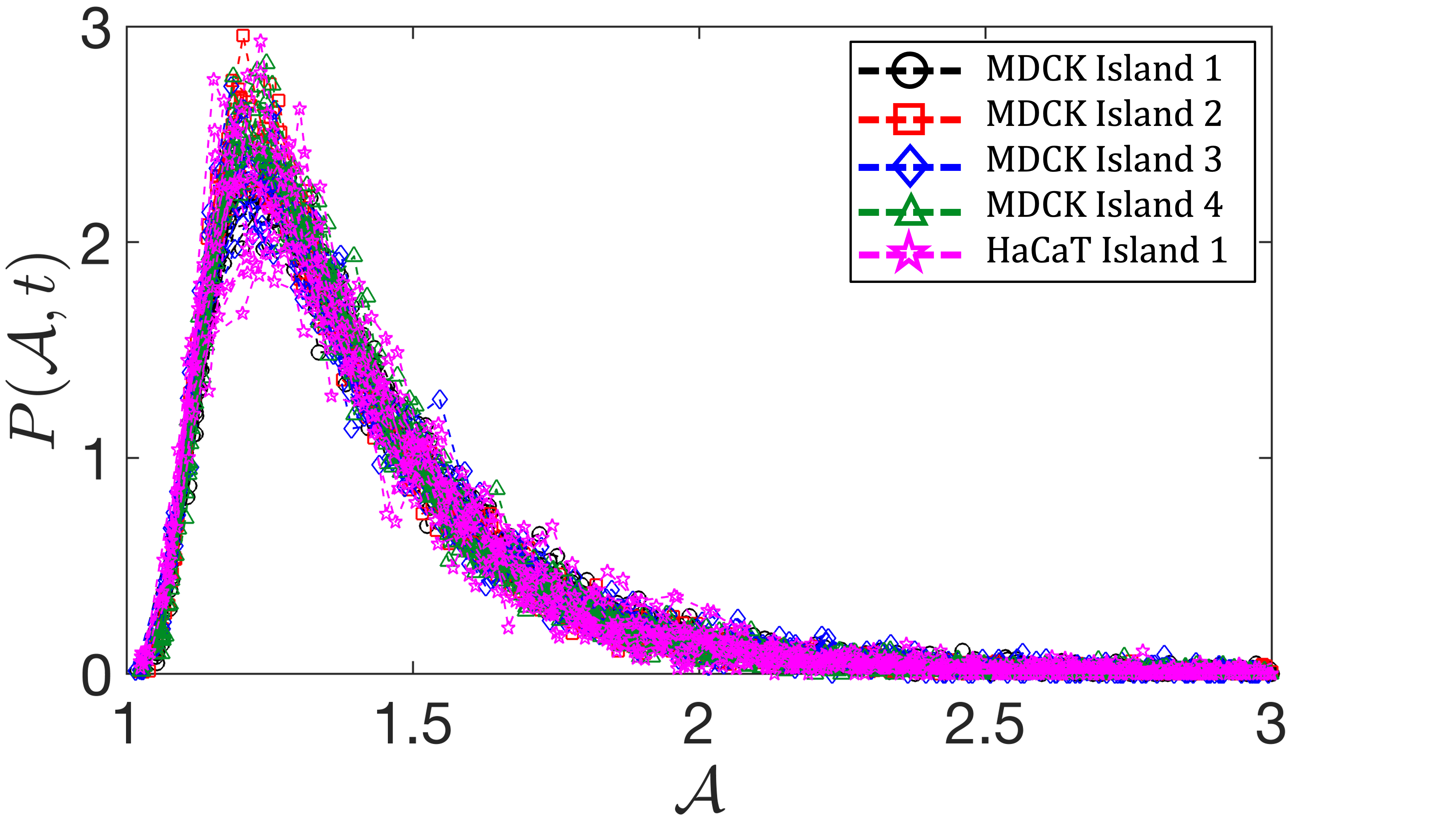}
    \caption{The shape parameter distributions $P(\calA, t)$ for four MDCK (circles, squares, diamonds, and triangles) and one HaCaT (stars) epithelial cell islands at $18$ equally-spaced time points $t$ separated by $10$ minutes.}
    \label{fig:shape_single_time}
\end{figure}

\begin{figure}[tb]

    \includegraphics[width=\columnwidth]{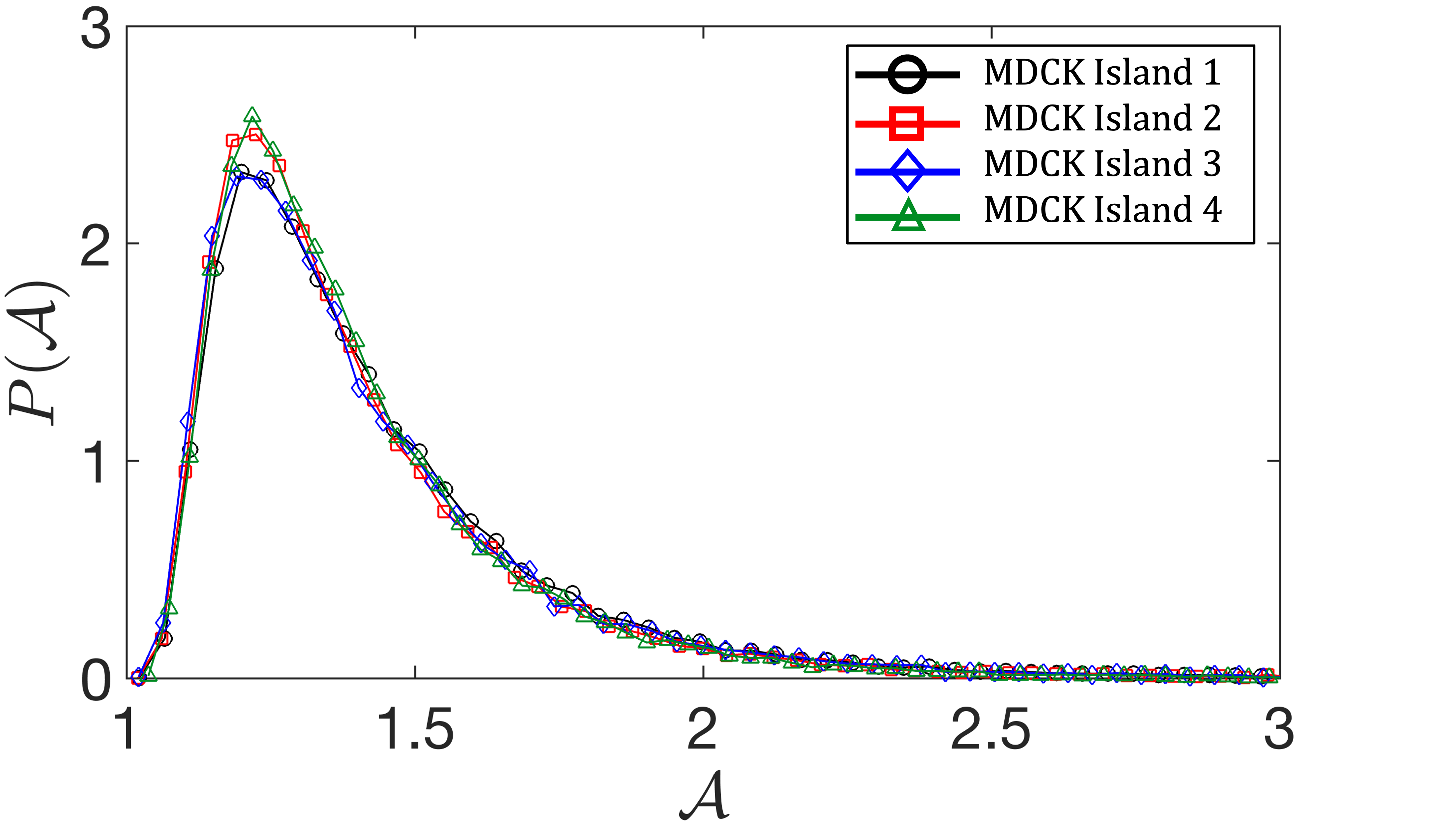}
    \caption{The shape parameter distributions for four low-density MDCK epithelial cell islands averaged over cells and time. This data is averaged to generate $P(\calA)$ for low-density MDCK epithelial cell islands in Fig.~\ref{fig:MDCK}.
    }
    \label{fig:single_islands}
\end{figure}

In Fig.~\ref{fig:MDCK} (b), we showed that the shape parameter distribution of individual epithelial cells as a function of time is similar to that for $P(\calA)$, binned over cells, times, and replicates. To assess the validity of averaging over different times and datasets, we show in  Fig.~\ref{fig:shape_single_time} that the shape parameter distributions $P(\calA, t)$ at each time and for different low-density islands of MDCK cells are similar.  Thus, the time- and spatially-averaged shape parameter distributions $P(\calA)$ for each of the low-density islands are also similar as shown in Fig.~\ref{fig:single_islands}. 

\section*{Appendix C: Fitting the Experimental Shape Parameter Distribution}

\begin{figure}[h!]
    \includegraphics[width=\columnwidth]{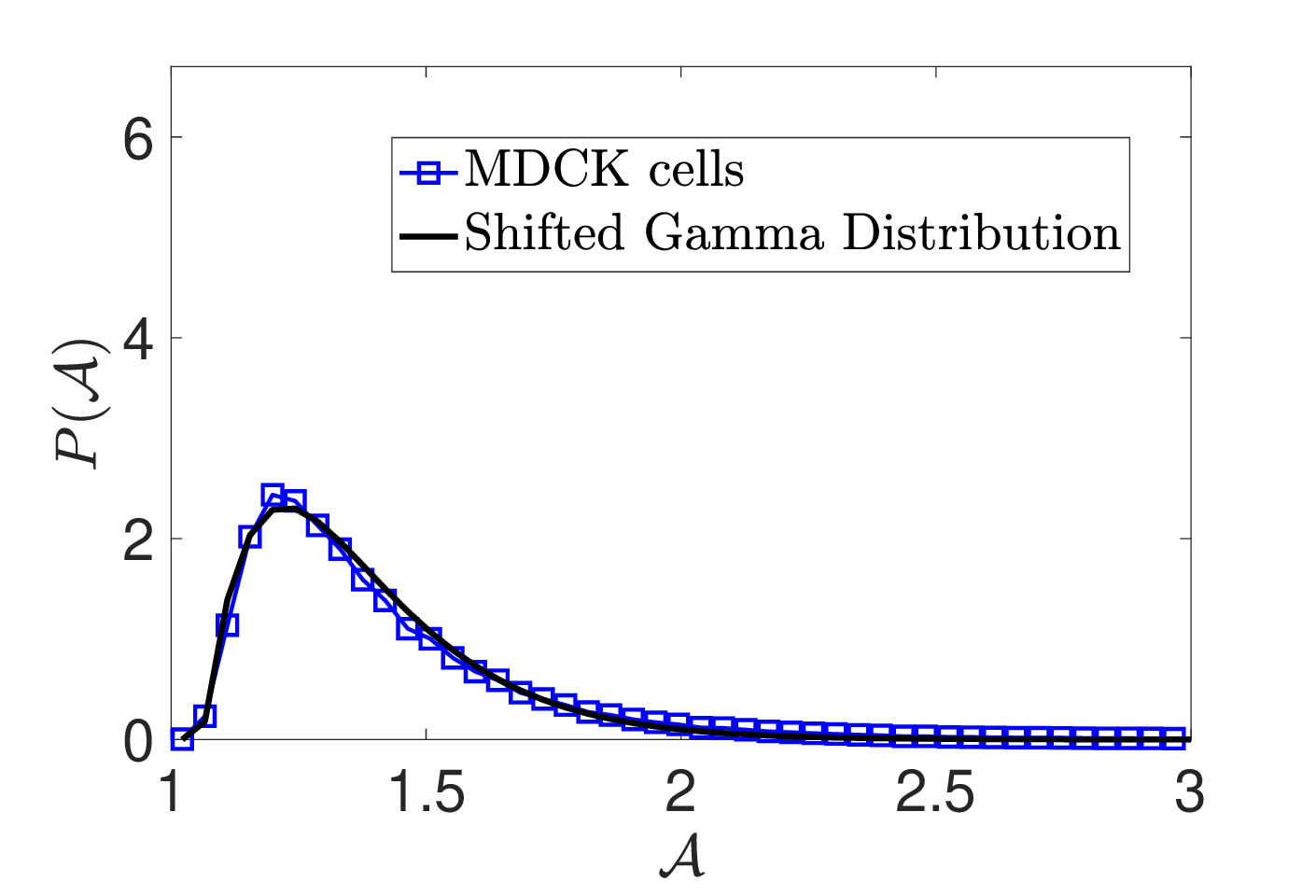}
    \caption{The shape parameter distribution $P({\cal A})$ (blue solid line) from low-density islands of MDCK epithelial cells in dataset (1). In addition, we show the fit of $P({\cal A})$ to a shifted gamma distribution (Eq. \ref{eq:gamma_fit}) with $k=2.0$, $\theta=0.16$, and $s=1.06$.}
    \label{fig:MDCK_gamma_fit}
\end{figure}

In this appendix, we fit the shape distribution for the low-density islands of epithelial cells to a shifted gamma distribution. A shifted gamma distribution is necessary, since the minimum shape parameter is $\calA=1.0$, not zero. The shifted gamma distribution is given by:
\begin{equation}
    P(\mathcal{A}) = \frac{1}{\Gamma(k)\theta^{k}}(\mathcal{A}-s)^{k-1}e^{-\frac{(\mathcal{A}-s)}{\theta}},
    \label{eq:gamma_fit}
\end{equation}
where $k$ and $\theta$ are the scale and shape parameters for the distribution, $\Gamma(k)$ is the gamma function, and $s$ is the shift parameter. In Fig. \ref{fig:MDCK_gamma_fit} we compare the best fit of $P({\cal A})$ for the low-density islands of MDCK epithelial cells to Eq.~\ref{eq:gamma_fit} with a normalized root-mean-squared-error of $2.5\%$.

\section*{Appendix D: Comparison of Shape Parameter Distributions of Low- and High-density Epithelial Cell Monolayers}

\label{sec:cell-line-shape-dists}

\begin{figure}[tb]

    \includegraphics[width=\columnwidth]{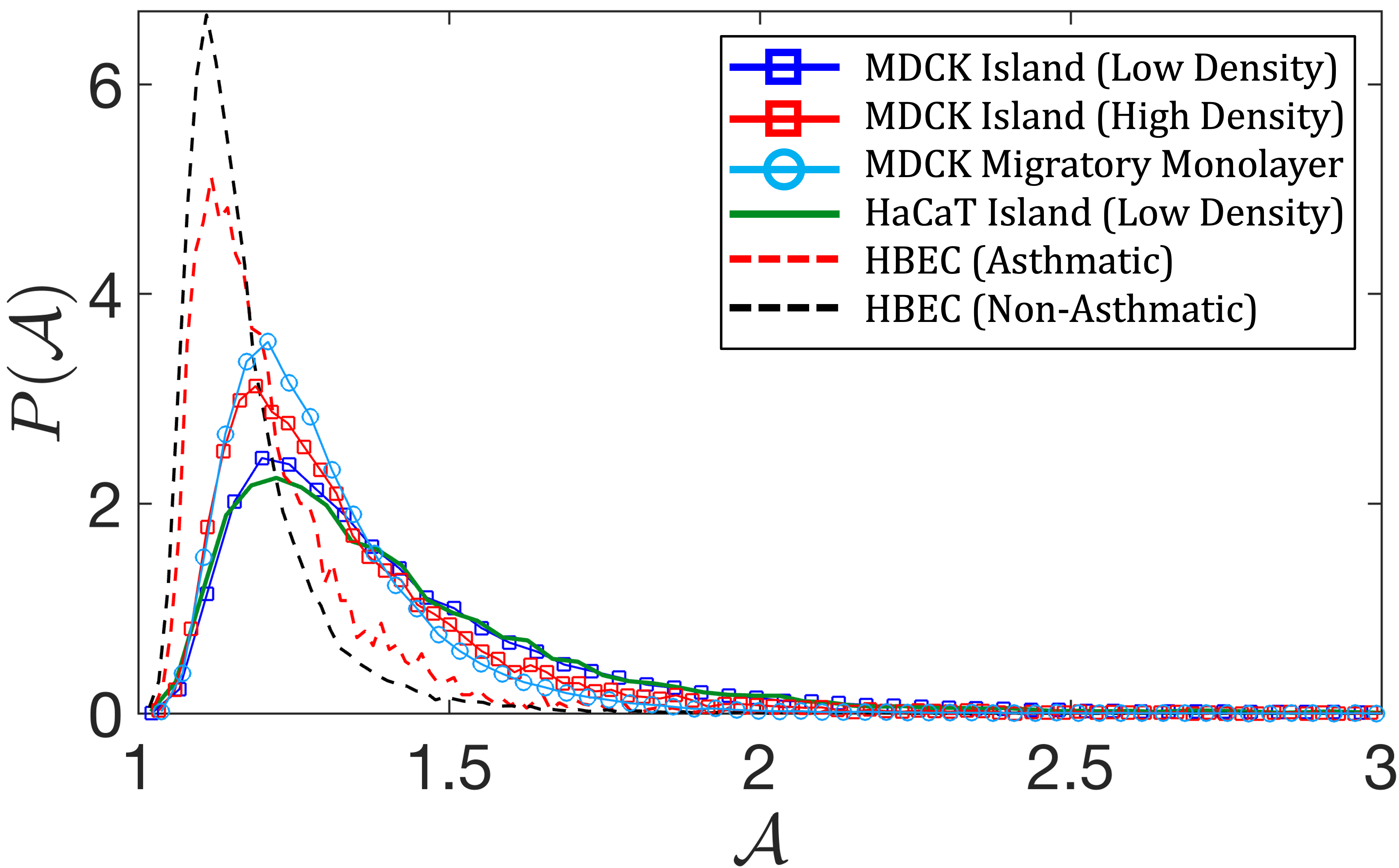}
    \caption{Shape parameter distributions $P(\calA)$ for the six epithelial cell monolayer datasets in Table~\ref{table:image_details}: (1) low- (blue squares) and (2) high-density  (red squares) islands of MDCK cells, (3) a single low-density island of HaCaT cells (green), (4) a single high-density MDCK migratory monolayer (sky-blue circles), and (5) asthmatic (red dashed) and (6) non-asthmatic (black dashed) human bronchial epithelial cell (HBEC) monolayers.
    }
    \label{fig:all-image-shape-dists}
\end{figure}

In this work, we measured the shape parameter distributions $P(\calA)$ of six types of epithelial cell monolayers: (1) low- and (2) high-density islands of MDCK cells, (3) a single low-density island of HaCaT cells, (4) a single high-density MDCK migratory monolayer, and chemically fixed (5) asthmatic and (6) non-asthmatic human bronchial (HB) epithelial cell monolayers. We focused on $P(\calA)$ for the low-density islands of MDCK and HaCaT epithelial cells since cells in these systems are more motile and undergo cell rearrangements with larger displacements compared to cells in high-density monolayers. We compare $P(\calA)$ fox the six types of epithelial cell monolayers in Fig.~\ref{fig:all-image-shape-dists}. We show that low-density, fluidized epithelial cell monolayers possess broad positively-skewed $P({\cal A})$, while high-density, more solid-like epithelial cell monolayers are strongly peaked with much smaller values for $\langle {\cal A} \rangle$, $\varcalA$, and $\skewcalA$.

\clearpage

\bibliography{references.bib}

\begin{thebibliography}{41}%
\makeatletter
\providecommand \@ifxundefined [1]{%
 \@ifx{#1\undefined}
}%
\providecommand \@ifnum [1]{%
 \ifnum #1\expandafter \@firstoftwo
 \else \expandafter \@secondoftwo
 \fi
}%
\providecommand \@ifx [1]{%
 \ifx #1\expandafter \@firstoftwo
 \else \expandafter \@secondoftwo
 \fi
}%
\providecommand \natexlab [1]{#1}%
\providecommand \enquote  [1]{``#1''}%
\providecommand \bibnamefont  [1]{#1}%
\providecommand \bibfnamefont [1]{#1}%
\providecommand \citenamefont [1]{#1}%
\providecommand \href@noop [0]{\@secondoftwo}%
\providecommand \href [0]{\begingroup \@sanitize@url \@href}%
\providecommand \@href[1]{\@@startlink{#1}\@@href}%
\providecommand \@@href[1]{\endgroup#1\@@endlink}%
\providecommand \@sanitize@url [0]{\catcode `\\12\catcode `\$12\catcode `\&12\catcode `\#12\catcode `\^12\catcode `\_12\catcode `\%12\relax}%
\providecommand \@@startlink[1]{}%
\providecommand \@@endlink[0]{}%
\providecommand \url  [0]{\begingroup\@sanitize@url \@url }%
\providecommand \@url [1]{\endgroup\@href {#1}{\urlprefix }}%
\providecommand \urlprefix  [0]{URL }%
\providecommand \Eprint [0]{\href }%
\providecommand \doibase [0]{https://doi.org/}%
\providecommand \selectlanguage [0]{\@gobble}%
\providecommand \bibinfo  [0]{\@secondoftwo}%
\providecommand \bibfield  [0]{\@secondoftwo}%
\providecommand \translation [1]{[#1]}%
\providecommand \BibitemOpen [0]{}%
\providecommand \bibitemStop [0]{}%
\providecommand \bibitemNoStop [0]{.\EOS\space}%
\providecommand \EOS [0]{\spacefactor3000\relax}%
\providecommand \BibitemShut  [1]{\csname bibitem#1\endcsname}%
\let\auto@bib@innerbib\@empty
\bibitem [{\citenamefont {Pusey}\ and\ \citenamefont {Van~Megen}(1986)}]{pusey1986phase}%
  \BibitemOpen
  \bibfield  {author} {\bibinfo {author} {\bibfnamefont {P.~N.}\ \bibnamefont {Pusey}}\ and\ \bibinfo {author} {\bibfnamefont {W.}~\bibnamefont {Van~Megen}},\ }\bibfield  {title} {\bibinfo {title} {Phase behaviour of concentrated suspensions of nearly hard colloidal spheres},\ }\href@noop {} {\bibfield  {journal} {\bibinfo  {journal} {Nature}\ }\textbf {\bibinfo {volume} {320}},\ \bibinfo {pages} {340} (\bibinfo {year} {1986})}\BibitemShut {NoStop}%
\bibitem [{\citenamefont {Behringer}\ and\ \citenamefont {Chakraborty}(2018)}]{behringer2018physics}%
  \BibitemOpen
  \bibfield  {author} {\bibinfo {author} {\bibfnamefont {R.~P.}\ \bibnamefont {Behringer}}\ and\ \bibinfo {author} {\bibfnamefont {B.}~\bibnamefont {Chakraborty}},\ }\bibfield  {title} {\bibinfo {title} {The physics of jamming for granular materials: a review},\ }\href@noop {} {\bibfield  {journal} {\bibinfo  {journal} {Reports on Progress in Physics}\ }\textbf {\bibinfo {volume} {82}},\ \bibinfo {pages} {012601} (\bibinfo {year} {2018})}\BibitemShut {NoStop}%
\bibitem [{\citenamefont {Pusey}\ and\ \citenamefont {van Megen}(1987)}]{PhysRevLett.59.2083}%
  \BibitemOpen
  \bibfield  {author} {\bibinfo {author} {\bibfnamefont {P.~N.}\ \bibnamefont {Pusey}}\ and\ \bibinfo {author} {\bibfnamefont {W.}~\bibnamefont {van Megen}},\ }\bibfield  {title} {\bibinfo {title} {Observation of a glass transition in suspensions of spherical colloidal particles},\ }\href {https://doi.org/10.1103/PhysRevLett.59.2083} {\bibfield  {journal} {\bibinfo  {journal} {Phys. Rev. Lett.}\ }\textbf {\bibinfo {volume} {59}},\ \bibinfo {pages} {2083} (\bibinfo {year} {1987})}\BibitemShut {NoStop}%
\bibitem [{\citenamefont {Doostmohammadi}\ \emph {et~al.}(2015)\citenamefont {Doostmohammadi}, \citenamefont {Thampi}, \citenamefont {Saw}, \citenamefont {Lim}, \citenamefont {Ladoux},\ and\ \citenamefont {Yeomans}}]{doostmohammadi2015celebrating}%
  \BibitemOpen
  \bibfield  {author} {\bibinfo {author} {\bibfnamefont {A.}~\bibnamefont {Doostmohammadi}}, \bibinfo {author} {\bibfnamefont {S.~P.}\ \bibnamefont {Thampi}}, \bibinfo {author} {\bibfnamefont {T.~B.}\ \bibnamefont {Saw}}, \bibinfo {author} {\bibfnamefont {C.~T.}\ \bibnamefont {Lim}}, \bibinfo {author} {\bibfnamefont {B.}~\bibnamefont {Ladoux}},\ and\ \bibinfo {author} {\bibfnamefont {J.~M.}\ \bibnamefont {Yeomans}},\ }\bibfield  {title} {\bibinfo {title} {Celebrating soft matter's 10th anniversary: Cell division: a source of active stress in cellular monolayers},\ }\href@noop {} {\bibfield  {journal} {\bibinfo  {journal} {Soft Matter}\ }\textbf {\bibinfo {volume} {11}},\ \bibinfo {pages} {7328} (\bibinfo {year} {2015})}\BibitemShut {NoStop}%
\bibitem [{\citenamefont {Saraswathibhatla}\ and\ \citenamefont {Notbohm}(2020)}]{saraswathibhatla2020tractions}%
  \BibitemOpen
  \bibfield  {author} {\bibinfo {author} {\bibfnamefont {A.}~\bibnamefont {Saraswathibhatla}}\ and\ \bibinfo {author} {\bibfnamefont {J.}~\bibnamefont {Notbohm}},\ }\bibfield  {title} {\bibinfo {title} {Tractions and stress fibers control cell shape and rearrangements in collective cell migration},\ }\href@noop {} {\bibfield  {journal} {\bibinfo  {journal} {Physical Review X}\ }\textbf {\bibinfo {volume} {10}},\ \bibinfo {pages} {011016} (\bibinfo {year} {2020})}\BibitemShut {NoStop}%
\bibitem [{\citenamefont {Atia}\ \emph {et~al.}(2018)\citenamefont {Atia}, \citenamefont {Bi}, \citenamefont {Sharma}, \citenamefont {Mitchel}, \citenamefont {Gweon}, \citenamefont {A.~Koehler}, \citenamefont {DeCamp}, \citenamefont {Lan}, \citenamefont {Kim}, \citenamefont {Hirsch} \emph {et~al.}}]{atia2018geometric}%
  \BibitemOpen
  \bibfield  {author} {\bibinfo {author} {\bibfnamefont {L.}~\bibnamefont {Atia}}, \bibinfo {author} {\bibfnamefont {D.}~\bibnamefont {Bi}}, \bibinfo {author} {\bibfnamefont {Y.}~\bibnamefont {Sharma}}, \bibinfo {author} {\bibfnamefont {J.~A.}\ \bibnamefont {Mitchel}}, \bibinfo {author} {\bibfnamefont {B.}~\bibnamefont {Gweon}}, \bibinfo {author} {\bibfnamefont {S.}~\bibnamefont {A.~Koehler}}, \bibinfo {author} {\bibfnamefont {S.~J.}\ \bibnamefont {DeCamp}}, \bibinfo {author} {\bibfnamefont {B.}~\bibnamefont {Lan}}, \bibinfo {author} {\bibfnamefont {J.~H.}\ \bibnamefont {Kim}}, \bibinfo {author} {\bibfnamefont {R.}~\bibnamefont {Hirsch}}, \emph {et~al.},\ }\bibfield  {title} {\bibinfo {title} {Geometric constraints during epithelial jamming},\ }\href@noop {} {\bibfield  {journal} {\bibinfo  {journal} {Nature physics}\ }\textbf {\bibinfo {volume} {14}},\ \bibinfo {pages} {613} (\bibinfo {year} {2018})}\BibitemShut {NoStop}%
\bibitem [{\citenamefont {Boromand}\ \emph {et~al.}(2018)\citenamefont {Boromand}, \citenamefont {Signoriello}, \citenamefont {Ye}, \citenamefont {O’Hern},\ and\ \citenamefont {Shattuck}}]{boromand2018jamming}%
  \BibitemOpen
  \bibfield  {author} {\bibinfo {author} {\bibfnamefont {A.}~\bibnamefont {Boromand}}, \bibinfo {author} {\bibfnamefont {A.}~\bibnamefont {Signoriello}}, \bibinfo {author} {\bibfnamefont {F.}~\bibnamefont {Ye}}, \bibinfo {author} {\bibfnamefont {C.~S.}\ \bibnamefont {O’Hern}},\ and\ \bibinfo {author} {\bibfnamefont {M.~D.}\ \bibnamefont {Shattuck}},\ }\bibfield  {title} {\bibinfo {title} {Jamming of deformable polygons},\ }\href@noop {} {\bibfield  {journal} {\bibinfo  {journal} {Physical review letters}\ }\textbf {\bibinfo {volume} {m121}},\ \bibinfo {pages} {248003} (\bibinfo {year} {2018})}\BibitemShut {NoStop}%
\bibitem [{\citenamefont {Fern{\'a}ndez}\ and\ \citenamefont {Ott}(2008)}]{fernandez2008single}%
  \BibitemOpen
  \bibfield  {author} {\bibinfo {author} {\bibfnamefont {P.}~\bibnamefont {Fern{\'a}ndez}}\ and\ \bibinfo {author} {\bibfnamefont {A.}~\bibnamefont {Ott}},\ }\bibfield  {title} {\bibinfo {title} {Single cell mechanics: stress stiffening and kinematic hardening},\ }\href@noop {} {\bibfield  {journal} {\bibinfo  {journal} {Physical Review Letters}\ }\textbf {\bibinfo {volume} {100}},\ \bibinfo {pages} {238102} (\bibinfo {year} {2008})}\BibitemShut {NoStop}%
\bibitem [{\citenamefont {Cordes}\ \emph {et~al.}(2020)\citenamefont {Cordes}, \citenamefont {Witt}, \citenamefont {Gallem{\'\i}-P{\'e}rez}, \citenamefont {Br{\"u}ckner}, \citenamefont {Grimm}, \citenamefont {Vache}, \citenamefont {Oswald}, \citenamefont {Bodenschatz}, \citenamefont {Flormann}, \citenamefont {Lautenschl{\"a}ger} \emph {et~al.}}]{cordes2020prestress}%
  \BibitemOpen
  \bibfield  {author} {\bibinfo {author} {\bibfnamefont {A.}~\bibnamefont {Cordes}}, \bibinfo {author} {\bibfnamefont {H.}~\bibnamefont {Witt}}, \bibinfo {author} {\bibfnamefont {A.}~\bibnamefont {Gallem{\'\i}-P{\'e}rez}}, \bibinfo {author} {\bibfnamefont {B.}~\bibnamefont {Br{\"u}ckner}}, \bibinfo {author} {\bibfnamefont {F.}~\bibnamefont {Grimm}}, \bibinfo {author} {\bibfnamefont {M.}~\bibnamefont {Vache}}, \bibinfo {author} {\bibfnamefont {T.}~\bibnamefont {Oswald}}, \bibinfo {author} {\bibfnamefont {J.}~\bibnamefont {Bodenschatz}}, \bibinfo {author} {\bibfnamefont {D.}~\bibnamefont {Flormann}}, \bibinfo {author} {\bibfnamefont {F.}~\bibnamefont {Lautenschl{\"a}ger}}, \emph {et~al.},\ }\bibfield  {title} {\bibinfo {title} {Prestress and area compressibility of actin cortices determine the viscoelastic response of living cells},\ }\href@noop {} {\bibfield  {journal} {\bibinfo  {journal} {Physical Review Letters}\ }\textbf {\bibinfo {volume} {125}},\ \bibinfo {pages} {068101} (\bibinfo {year}
  {2020})}\BibitemShut {NoStop}%
\bibitem [{\citenamefont {Chugh}\ and\ \citenamefont {Paluch}(2018)}]{chugh2018actin}%
  \BibitemOpen
  \bibfield  {author} {\bibinfo {author} {\bibfnamefont {P.}~\bibnamefont {Chugh}}\ and\ \bibinfo {author} {\bibfnamefont {E.~K.}\ \bibnamefont {Paluch}},\ }\bibfield  {title} {\bibinfo {title} {The actin cortex at a glance},\ }\href@noop {} {\bibfield  {journal} {\bibinfo  {journal} {Journal of cell science}\ }\textbf {\bibinfo {volume} {131}},\ \bibinfo {pages} {jcs186254} (\bibinfo {year} {2018})}\BibitemShut {NoStop}%
\bibitem [{\citenamefont {Bi}\ \emph {et~al.}(2015)\citenamefont {Bi}, \citenamefont {Lopez}, \citenamefont {Schwarz},\ and\ \citenamefont {Manning}}]{bi2015density}%
  \BibitemOpen
  \bibfield  {author} {\bibinfo {author} {\bibfnamefont {D.}~\bibnamefont {Bi}}, \bibinfo {author} {\bibfnamefont {J.}~\bibnamefont {Lopez}}, \bibinfo {author} {\bibfnamefont {J.~M.}\ \bibnamefont {Schwarz}},\ and\ \bibinfo {author} {\bibfnamefont {M.~L.}\ \bibnamefont {Manning}},\ }\bibfield  {title} {\bibinfo {title} {A density-independent rigidity transition in biological tissues},\ }\href@noop {} {\bibfield  {journal} {\bibinfo  {journal} {Nature Physics}\ }\textbf {\bibinfo {volume} {11}},\ \bibinfo {pages} {1074} (\bibinfo {year} {2015})}\BibitemShut {NoStop}%
\bibitem [{\citenamefont {Bi}\ \emph {et~al.}(2016)\citenamefont {Bi}, \citenamefont {Yang}, \citenamefont {Marchetti},\ and\ \citenamefont {Manning}}]{bi2016voronoi}%
  \BibitemOpen
  \bibfield  {author} {\bibinfo {author} {\bibfnamefont {D.}~\bibnamefont {Bi}}, \bibinfo {author} {\bibfnamefont {X.}~\bibnamefont {Yang}}, \bibinfo {author} {\bibfnamefont {M.~C.}\ \bibnamefont {Marchetti}},\ and\ \bibinfo {author} {\bibfnamefont {M.~L.}\ \bibnamefont {Manning}},\ }\bibfield  {title} {\bibinfo {title} {Motility-driven glass and jamming transitions in biological tissues},\ }\href@noop {} {\bibfield  {journal} {\bibinfo  {journal} {Physical Review X}\ }\textbf {\bibinfo {volume} {6}},\ \bibinfo {pages} {021011} (\bibinfo {year} {2016})}\BibitemShut {NoStop}%
\bibitem [{\citenamefont {Merkel}\ and\ \citenamefont {Manning}(2018)}]{merkel2018geometrically}%
  \BibitemOpen
  \bibfield  {author} {\bibinfo {author} {\bibfnamefont {M.}~\bibnamefont {Merkel}}\ and\ \bibinfo {author} {\bibfnamefont {M.~L.}\ \bibnamefont {Manning}},\ }\bibfield  {title} {\bibinfo {title} {A geometrically controlled rigidity transition in a model for confluent 3d tissues},\ }\href@noop {} {\bibfield  {journal} {\bibinfo  {journal} {New Journal of Physics}\ }\textbf {\bibinfo {volume} {20}},\ \bibinfo {pages} {022002} (\bibinfo {year} {2018})}\BibitemShut {NoStop}%
\bibitem [{\citenamefont {Sadhukhan}\ and\ \citenamefont {Nandi}(2022)}]{sadhukhan2022origin}%
  \BibitemOpen
  \bibfield  {author} {\bibinfo {author} {\bibfnamefont {S.}~\bibnamefont {Sadhukhan}}\ and\ \bibinfo {author} {\bibfnamefont {S.~K.}\ \bibnamefont {Nandi}},\ }\bibfield  {title} {\bibinfo {title} {On the origin of universal cell shape variability in confluent epithelial monolayers},\ }\href@noop {} {\bibfield  {journal} {\bibinfo  {journal} {Elife}\ }\textbf {\bibinfo {volume} {11}},\ \bibinfo {pages} {e76406} (\bibinfo {year} {2022})}\BibitemShut {NoStop}%
\bibitem [{\citenamefont {Saraswathibhatla}\ \emph {et~al.}(2020)\citenamefont {Saraswathibhatla}, \citenamefont {Galles},\ and\ \citenamefont {Notbohm}}]{saraswathibhatla2020spatiotemporal}%
  \BibitemOpen
  \bibfield  {author} {\bibinfo {author} {\bibfnamefont {A.}~\bibnamefont {Saraswathibhatla}}, \bibinfo {author} {\bibfnamefont {E.~E.}\ \bibnamefont {Galles}},\ and\ \bibinfo {author} {\bibfnamefont {J.}~\bibnamefont {Notbohm}},\ }\bibfield  {title} {\bibinfo {title} {Spatiotemporal force and motion in collective cell migration},\ }\href@noop {} {\bibfield  {journal} {\bibinfo  {journal} {Scientific Data}\ }\textbf {\bibinfo {volume} {7}},\ \bibinfo {pages} {197} (\bibinfo {year} {2020})}\BibitemShut {NoStop}%
\bibitem [{\citenamefont {Rosen}\ and\ \citenamefont {Misfeldt}(1980)}]{rosen1980cell}%
  \BibitemOpen
  \bibfield  {author} {\bibinfo {author} {\bibfnamefont {P.}~\bibnamefont {Rosen}}\ and\ \bibinfo {author} {\bibfnamefont {D.~S.}\ \bibnamefont {Misfeldt}},\ }\bibfield  {title} {\bibinfo {title} {Cell density determines epithelial migration in culture.},\ }\href@noop {} {\bibfield  {journal} {\bibinfo  {journal} {Proceedings of the National Academy of Sciences}\ }\textbf {\bibinfo {volume} {77}},\ \bibinfo {pages} {4760} (\bibinfo {year} {1980})}\BibitemShut {NoStop}%
\bibitem [{\citenamefont {Szabo}\ \emph {et~al.}(2001)\citenamefont {Szabo}, \citenamefont {Wetzel},\ and\ \citenamefont {Rogers}}]{szabo2001cell}%
  \BibitemOpen
  \bibfield  {author} {\bibinfo {author} {\bibfnamefont {I.}~\bibnamefont {Szabo}}, \bibinfo {author} {\bibfnamefont {M.~A.}\ \bibnamefont {Wetzel}},\ and\ \bibinfo {author} {\bibfnamefont {T.~J.}\ \bibnamefont {Rogers}},\ }\bibfield  {title} {\bibinfo {title} {Cell-density-regulated chemotactic responsiveness of keratinocytes in vitro},\ }\href@noop {} {\bibfield  {journal} {\bibinfo  {journal} {Journal of investigative dermatology}\ }\textbf {\bibinfo {volume} {117}},\ \bibinfo {pages} {1083} (\bibinfo {year} {2001})}\BibitemShut {NoStop}%
\bibitem [{\citenamefont {Garcia}\ \emph {et~al.}(2015)\citenamefont {Garcia}, \citenamefont {Hannezo}, \citenamefont {Elgeti}, \citenamefont {Joanny}, \citenamefont {Silberzan},\ and\ \citenamefont {Gov}}]{garcia2015physics}%
  \BibitemOpen
  \bibfield  {author} {\bibinfo {author} {\bibfnamefont {S.}~\bibnamefont {Garcia}}, \bibinfo {author} {\bibfnamefont {E.}~\bibnamefont {Hannezo}}, \bibinfo {author} {\bibfnamefont {J.}~\bibnamefont {Elgeti}}, \bibinfo {author} {\bibfnamefont {J.-F.}\ \bibnamefont {Joanny}}, \bibinfo {author} {\bibfnamefont {P.}~\bibnamefont {Silberzan}},\ and\ \bibinfo {author} {\bibfnamefont {N.~S.}\ \bibnamefont {Gov}},\ }\bibfield  {title} {\bibinfo {title} {Physics of active jamming during collective cellular motion in a monolayer},\ }\href@noop {} {\bibfield  {journal} {\bibinfo  {journal} {Proceedings of the National Academy of Sciences}\ }\textbf {\bibinfo {volume} {112}},\ \bibinfo {pages} {15314} (\bibinfo {year} {2015})}\BibitemShut {NoStop}%
\bibitem [{\citenamefont {Stringer}\ \emph {et~al.}(2021)\citenamefont {Stringer}, \citenamefont {Wang}, \citenamefont {Michaelos},\ and\ \citenamefont {Pachitariu}}]{stringer2021cellpose}%
  \BibitemOpen
  \bibfield  {author} {\bibinfo {author} {\bibfnamefont {C.}~\bibnamefont {Stringer}}, \bibinfo {author} {\bibfnamefont {T.}~\bibnamefont {Wang}}, \bibinfo {author} {\bibfnamefont {M.}~\bibnamefont {Michaelos}},\ and\ \bibinfo {author} {\bibfnamefont {M.}~\bibnamefont {Pachitariu}},\ }\bibfield  {title} {\bibinfo {title} {Cellpose: a generalist algorithm for cellular segmentation},\ }\href@noop {} {\bibfield  {journal} {\bibinfo  {journal} {Nature methods}\ }\textbf {\bibinfo {volume} {18}},\ \bibinfo {pages} {100} (\bibinfo {year} {2021})}\BibitemShut {NoStop}%
\bibitem [{\citenamefont {Pachitariu}\ and\ \citenamefont {Stringer}(2022)}]{pachitariu2022cellpose}%
  \BibitemOpen
  \bibfield  {author} {\bibinfo {author} {\bibfnamefont {M.}~\bibnamefont {Pachitariu}}\ and\ \bibinfo {author} {\bibfnamefont {C.}~\bibnamefont {Stringer}},\ }\bibfield  {title} {\bibinfo {title} {Cellpose 2.0: how to train your own model},\ }\href@noop {} {\bibfield  {journal} {\bibinfo  {journal} {Nature methods}\ }\textbf {\bibinfo {volume} {19}},\ \bibinfo {pages} {1634} (\bibinfo {year} {2022})}\BibitemShut {NoStop}%
\bibitem [{\citenamefont {Gerv{\'a}sio}\ \emph {et~al.}(2011)\citenamefont {Gerv{\'a}sio}, \citenamefont {Phillips}, \citenamefont {Cole},\ and\ \citenamefont {Allen}}]{gervasio2011caveolae}%
  \BibitemOpen
  \bibfield  {author} {\bibinfo {author} {\bibfnamefont {O.~L.}\ \bibnamefont {Gerv{\'a}sio}}, \bibinfo {author} {\bibfnamefont {W.~D.}\ \bibnamefont {Phillips}}, \bibinfo {author} {\bibfnamefont {L.}~\bibnamefont {Cole}},\ and\ \bibinfo {author} {\bibfnamefont {D.~G.}\ \bibnamefont {Allen}},\ }\bibfield  {title} {\bibinfo {title} {Caveolae respond to cell stretch and contribute to stretch-induced signaling},\ }\href@noop {} {\bibfield  {journal} {\bibinfo  {journal} {Journal of cell science}\ }\textbf {\bibinfo {volume} {124}},\ \bibinfo {pages} {3581} (\bibinfo {year} {2011})}\BibitemShut {NoStop}%
\bibitem [{\citenamefont {Apodaca}(2002)}]{apodaca2002modulation}%
  \BibitemOpen
  \bibfield  {author} {\bibinfo {author} {\bibfnamefont {G.}~\bibnamefont {Apodaca}},\ }\bibfield  {title} {\bibinfo {title} {Modulation of membrane traffic by mechanical stimuli},\ }\href@noop {} {\bibfield  {journal} {\bibinfo  {journal} {American Journal of Physiology-Renal Physiology}\ }\textbf {\bibinfo {volume} {282}},\ \bibinfo {pages} {F179} (\bibinfo {year} {2002})}\BibitemShut {NoStop}%
\bibitem [{\citenamefont {Sheetz}(2001)}]{sheetz2001cell}%
  \BibitemOpen
  \bibfield  {author} {\bibinfo {author} {\bibfnamefont {M.~P.}\ \bibnamefont {Sheetz}},\ }\bibfield  {title} {\bibinfo {title} {Cell control by membrane--cytoskeleton adhesion},\ }\href@noop {} {\bibfield  {journal} {\bibinfo  {journal} {Nature Reviews Molecular Cell Biology}\ }\textbf {\bibinfo {volume} {2}},\ \bibinfo {pages} {392} (\bibinfo {year} {2001})}\BibitemShut {NoStop}%
\bibitem [{\citenamefont {Zehnder}\ \emph {et~al.}(2015)\citenamefont {Zehnder}, \citenamefont {Suaris}, \citenamefont {Bellaire},\ and\ \citenamefont {Angelini}}]{zehnder2015cell}%
  \BibitemOpen
  \bibfield  {author} {\bibinfo {author} {\bibfnamefont {S.~M.}\ \bibnamefont {Zehnder}}, \bibinfo {author} {\bibfnamefont {M.}~\bibnamefont {Suaris}}, \bibinfo {author} {\bibfnamefont {M.~M.}\ \bibnamefont {Bellaire}},\ and\ \bibinfo {author} {\bibfnamefont {T.~E.}\ \bibnamefont {Angelini}},\ }\bibfield  {title} {\bibinfo {title} {Cell volume fluctuations in mdck monolayers},\ }\href@noop {} {\bibfield  {journal} {\bibinfo  {journal} {Biophysical journal}\ }\textbf {\bibinfo {volume} {108}},\ \bibinfo {pages} {247} (\bibinfo {year} {2015})}\BibitemShut {NoStop}%
\bibitem [{\citenamefont {Roshal}\ \emph {et~al.}(2022)\citenamefont {Roshal}, \citenamefont {Martin}, \citenamefont {Fedorenko}, \citenamefont {Golushko}, \citenamefont {Molle}, \citenamefont {Baghdiguian},\ and\ \citenamefont {Rochal}}]{roshal2022random}%
  \BibitemOpen
  \bibfield  {author} {\bibinfo {author} {\bibfnamefont {D.~S.}\ \bibnamefont {Roshal}}, \bibinfo {author} {\bibfnamefont {M.}~\bibnamefont {Martin}}, \bibinfo {author} {\bibfnamefont {K.}~\bibnamefont {Fedorenko}}, \bibinfo {author} {\bibfnamefont {I.}~\bibnamefont {Golushko}}, \bibinfo {author} {\bibfnamefont {V.}~\bibnamefont {Molle}}, \bibinfo {author} {\bibfnamefont {S.}~\bibnamefont {Baghdiguian}},\ and\ \bibinfo {author} {\bibfnamefont {S.~B.}\ \bibnamefont {Rochal}},\ }\bibfield  {title} {\bibinfo {title} {Random nature of epithelial cancer cell monolayers},\ }\href@noop {} {\bibfield  {journal} {\bibinfo  {journal} {Journal of the Royal Society Interface}\ }\textbf {\bibinfo {volume} {19}},\ \bibinfo {pages} {20220026} (\bibinfo {year} {2022})}\BibitemShut {NoStop}%
\bibitem [{\citenamefont {Ershov}\ \emph {et~al.}(2022)\citenamefont {Ershov}, \citenamefont {Phan}, \citenamefont {Pylv{\"a}n{\"a}inen}, \citenamefont {Rigaud}, \citenamefont {Le~Blanc}, \citenamefont {Charles-Orszag}, \citenamefont {Conway}, \citenamefont {Laine}, \citenamefont {Roy}, \citenamefont {Bonazzi} \emph {et~al.}}]{ershov2022trackmate}%
  \BibitemOpen
  \bibfield  {author} {\bibinfo {author} {\bibfnamefont {D.}~\bibnamefont {Ershov}}, \bibinfo {author} {\bibfnamefont {M.-S.}\ \bibnamefont {Phan}}, \bibinfo {author} {\bibfnamefont {J.~W.}\ \bibnamefont {Pylv{\"a}n{\"a}inen}}, \bibinfo {author} {\bibfnamefont {S.~U.}\ \bibnamefont {Rigaud}}, \bibinfo {author} {\bibfnamefont {L.}~\bibnamefont {Le~Blanc}}, \bibinfo {author} {\bibfnamefont {A.}~\bibnamefont {Charles-Orszag}}, \bibinfo {author} {\bibfnamefont {J.~R.}\ \bibnamefont {Conway}}, \bibinfo {author} {\bibfnamefont {R.~F.}\ \bibnamefont {Laine}}, \bibinfo {author} {\bibfnamefont {N.~H.}\ \bibnamefont {Roy}}, \bibinfo {author} {\bibfnamefont {D.}~\bibnamefont {Bonazzi}}, \emph {et~al.},\ }\bibfield  {title} {\bibinfo {title} {Trackmate 7: integrating state-of-the-art segmentation algorithms into tracking pipelines},\ }\href@noop {} {\bibfield  {journal} {\bibinfo  {journal} {Nature methods}\ }\textbf {\bibinfo {volume} {19}},\ \bibinfo {pages} {829} (\bibinfo {year} {2022})}\BibitemShut {NoStop}%
\bibitem [{\citenamefont {Tinevez}\ \emph {et~al.}(2017)\citenamefont {Tinevez}, \citenamefont {Perry}, \citenamefont {Schindelin}, \citenamefont {Hoopes}, \citenamefont {Reynolds}, \citenamefont {Laplantine}, \citenamefont {Bednarek}, \citenamefont {Shorte},\ and\ \citenamefont {Eliceiri}}]{tinevez2017trackmate}%
  \BibitemOpen
  \bibfield  {author} {\bibinfo {author} {\bibfnamefont {J.-Y.}\ \bibnamefont {Tinevez}}, \bibinfo {author} {\bibfnamefont {N.}~\bibnamefont {Perry}}, \bibinfo {author} {\bibfnamefont {J.}~\bibnamefont {Schindelin}}, \bibinfo {author} {\bibfnamefont {G.~M.}\ \bibnamefont {Hoopes}}, \bibinfo {author} {\bibfnamefont {G.~D.}\ \bibnamefont {Reynolds}}, \bibinfo {author} {\bibfnamefont {E.}~\bibnamefont {Laplantine}}, \bibinfo {author} {\bibfnamefont {S.~Y.}\ \bibnamefont {Bednarek}}, \bibinfo {author} {\bibfnamefont {S.~L.}\ \bibnamefont {Shorte}},\ and\ \bibinfo {author} {\bibfnamefont {K.~W.}\ \bibnamefont {Eliceiri}},\ }\bibfield  {title} {\bibinfo {title} {Trackmate: An open and extensible platform for single-particle tracking},\ }\href@noop {} {\bibfield  {journal} {\bibinfo  {journal} {Methods}\ }\textbf {\bibinfo {volume} {115}},\ \bibinfo {pages} {80} (\bibinfo {year} {2017})}\BibitemShut {NoStop}%
\bibitem [{\citenamefont {Nagle}\ \emph {et~al.}(2015)\citenamefont {Nagle}, \citenamefont {Jablin}, \citenamefont {Tristram-Nagle},\ and\ \citenamefont {Akabori}}]{nagle2015true}%
  \BibitemOpen
  \bibfield  {author} {\bibinfo {author} {\bibfnamefont {J.~F.}\ \bibnamefont {Nagle}}, \bibinfo {author} {\bibfnamefont {M.~S.}\ \bibnamefont {Jablin}}, \bibinfo {author} {\bibfnamefont {S.}~\bibnamefont {Tristram-Nagle}},\ and\ \bibinfo {author} {\bibfnamefont {K.}~\bibnamefont {Akabori}},\ }\bibfield  {title} {\bibinfo {title} {What are the true values of the bending modulus of simple lipid bilayers?},\ }\href@noop {} {\bibfield  {journal} {\bibinfo  {journal} {Chemistry and physics of lipids}\ }\textbf {\bibinfo {volume} {185}},\ \bibinfo {pages} {3} (\bibinfo {year} {2015})}\BibitemShut {NoStop}%
\bibitem [{\citenamefont {Treado}\ \emph {et~al.}(2021)\citenamefont {Treado}, \citenamefont {Wang}, \citenamefont {Boromand}, \citenamefont {Murrell}, \citenamefont {Shattuck},\ and\ \citenamefont {O'Hern}}]{treado2021bridging}%
  \BibitemOpen
  \bibfield  {author} {\bibinfo {author} {\bibfnamefont {J.~D.}\ \bibnamefont {Treado}}, \bibinfo {author} {\bibfnamefont {D.}~\bibnamefont {Wang}}, \bibinfo {author} {\bibfnamefont {A.}~\bibnamefont {Boromand}}, \bibinfo {author} {\bibfnamefont {M.~P.}\ \bibnamefont {Murrell}}, \bibinfo {author} {\bibfnamefont {M.~D.}\ \bibnamefont {Shattuck}},\ and\ \bibinfo {author} {\bibfnamefont {C.~S.}\ \bibnamefont {O'Hern}},\ }\bibfield  {title} {\bibinfo {title} {Bridging particle deformability and collective response in soft solids},\ }\href@noop {} {\bibfield  {journal} {\bibinfo  {journal} {Physical Review Materials}\ }\textbf {\bibinfo {volume} {5}},\ \bibinfo {pages} {055605} (\bibinfo {year} {2021})}\BibitemShut {NoStop}%
\bibitem [{\citenamefont {Br{\"u}nger}\ \emph {et~al.}(1984)\citenamefont {Br{\"u}nger}, \citenamefont {Brooks~III},\ and\ \citenamefont {Karplus}}]{brunger1984stochastic}%
  \BibitemOpen
  \bibfield  {author} {\bibinfo {author} {\bibfnamefont {A.}~\bibnamefont {Br{\"u}nger}}, \bibinfo {author} {\bibfnamefont {C.~L.}\ \bibnamefont {Brooks~III}},\ and\ \bibinfo {author} {\bibfnamefont {M.}~\bibnamefont {Karplus}},\ }\bibfield  {title} {\bibinfo {title} {Stochastic boundary conditions for molecular dynamics simulations of st2 water},\ }\href@noop {} {\bibfield  {journal} {\bibinfo  {journal} {Chemical physics letters}\ }\textbf {\bibinfo {volume} {105}},\ \bibinfo {pages} {495} (\bibinfo {year} {1984})}\BibitemShut {NoStop}%
\bibitem [{\citenamefont {Vanden-Eijnden}\ and\ \citenamefont {Ciccotti}(2006)}]{vanden2006second}%
  \BibitemOpen
  \bibfield  {author} {\bibinfo {author} {\bibfnamefont {E.}~\bibnamefont {Vanden-Eijnden}}\ and\ \bibinfo {author} {\bibfnamefont {G.}~\bibnamefont {Ciccotti}},\ }\bibfield  {title} {\bibinfo {title} {Second-order integrators for langevin equations with holonomic constraints},\ }\href@noop {} {\bibfield  {journal} {\bibinfo  {journal} {Chemical physics letters}\ }\textbf {\bibinfo {volume} {429}},\ \bibinfo {pages} {310} (\bibinfo {year} {2006})}\BibitemShut {NoStop}%
\bibitem [{\citenamefont {Nagle}(2013)}]{nagle2013introductory}%
  \BibitemOpen
  \bibfield  {author} {\bibinfo {author} {\bibfnamefont {J.~F.}\ \bibnamefont {Nagle}},\ }\bibfield  {title} {\bibinfo {title} {Introductory lecture: basic quantities in model biomembranes},\ }\href@noop {} {\bibfield  {journal} {\bibinfo  {journal} {Faraday discussions}\ }\textbf {\bibinfo {volume} {161}},\ \bibinfo {pages} {11} (\bibinfo {year} {2013})}\BibitemShut {NoStop}%
\bibitem [{\citenamefont {Peukes}\ and\ \citenamefont {Betz}(2014)}]{peukes2014direct}%
  \BibitemOpen
  \bibfield  {author} {\bibinfo {author} {\bibfnamefont {J.}~\bibnamefont {Peukes}}\ and\ \bibinfo {author} {\bibfnamefont {T.}~\bibnamefont {Betz}},\ }\bibfield  {title} {\bibinfo {title} {Direct measurement of the cortical tension during the growth of membrane blebs},\ }\href@noop {} {\bibfield  {journal} {\bibinfo  {journal} {Biophysical journal}\ }\textbf {\bibinfo {volume} {107}},\ \bibinfo {pages} {1810} (\bibinfo {year} {2014})}\BibitemShut {NoStop}%
\bibitem [{\citenamefont {Moazzeni}\ \emph {et~al.}(2021)\citenamefont {Moazzeni}, \citenamefont {Demiryurek}, \citenamefont {Yu}, \citenamefont {Shreiber}, \citenamefont {Zahn}, \citenamefont {Shan}, \citenamefont {Foty}, \citenamefont {Liu},\ and\ \citenamefont {Lin}}]{moazzeni2021single}%
  \BibitemOpen
  \bibfield  {author} {\bibinfo {author} {\bibfnamefont {S.}~\bibnamefont {Moazzeni}}, \bibinfo {author} {\bibfnamefont {Y.}~\bibnamefont {Demiryurek}}, \bibinfo {author} {\bibfnamefont {M.}~\bibnamefont {Yu}}, \bibinfo {author} {\bibfnamefont {D.~I.}\ \bibnamefont {Shreiber}}, \bibinfo {author} {\bibfnamefont {J.~D.}\ \bibnamefont {Zahn}}, \bibinfo {author} {\bibfnamefont {J.~W.}\ \bibnamefont {Shan}}, \bibinfo {author} {\bibfnamefont {R.~A.}\ \bibnamefont {Foty}}, \bibinfo {author} {\bibfnamefont {L.}~\bibnamefont {Liu}},\ and\ \bibinfo {author} {\bibfnamefont {H.}~\bibnamefont {Lin}},\ }\bibfield  {title} {\bibinfo {title} {Single-cell mechanical analysis and tension quantification via electrodeformation relaxation},\ }\href@noop {} {\bibfield  {journal} {\bibinfo  {journal} {Physical Review E}\ }\textbf {\bibinfo {volume} {103}},\ \bibinfo {pages} {032409} (\bibinfo {year} {2021})}\BibitemShut {NoStop}%
\bibitem [{\citenamefont {Dai}\ \emph {et~al.}(1999)\citenamefont {Dai}, \citenamefont {Ting-Beall}, \citenamefont {Hochmuth}, \citenamefont {Sheetz},\ and\ \citenamefont {Titus}}]{dai1999myosin}%
  \BibitemOpen
  \bibfield  {author} {\bibinfo {author} {\bibfnamefont {J.}~\bibnamefont {Dai}}, \bibinfo {author} {\bibfnamefont {H.~P.}\ \bibnamefont {Ting-Beall}}, \bibinfo {author} {\bibfnamefont {R.~M.}\ \bibnamefont {Hochmuth}}, \bibinfo {author} {\bibfnamefont {M.~P.}\ \bibnamefont {Sheetz}},\ and\ \bibinfo {author} {\bibfnamefont {M.~A.}\ \bibnamefont {Titus}},\ }\bibfield  {title} {\bibinfo {title} {Myosin i contributes to the generation of resting cortical tension},\ }\href@noop {} {\bibfield  {journal} {\bibinfo  {journal} {Biophysical journal}\ }\textbf {\bibinfo {volume} {77}},\ \bibinfo {pages} {1168} (\bibinfo {year} {1999})}\BibitemShut {NoStop}%
\bibitem [{\citenamefont {Angelini}\ \emph {et~al.}(2011)\citenamefont {Angelini}, \citenamefont {Hannezo}, \citenamefont {Trepat}, \citenamefont {Marquez}, \citenamefont {Fredberg},\ and\ \citenamefont {Weitz}}]{angelini2011glass}%
  \BibitemOpen
  \bibfield  {author} {\bibinfo {author} {\bibfnamefont {T.~E.}\ \bibnamefont {Angelini}}, \bibinfo {author} {\bibfnamefont {E.}~\bibnamefont {Hannezo}}, \bibinfo {author} {\bibfnamefont {X.}~\bibnamefont {Trepat}}, \bibinfo {author} {\bibfnamefont {M.}~\bibnamefont {Marquez}}, \bibinfo {author} {\bibfnamefont {J.~J.}\ \bibnamefont {Fredberg}},\ and\ \bibinfo {author} {\bibfnamefont {D.~A.}\ \bibnamefont {Weitz}},\ }\bibfield  {title} {\bibinfo {title} {Glass-like dynamics of collective cell migration},\ }\href@noop {} {\bibfield  {journal} {\bibinfo  {journal} {Proceedings of the National Academy of Sciences}\ }\textbf {\bibinfo {volume} {108}},\ \bibinfo {pages} {4714} (\bibinfo {year} {2011})}\BibitemShut {NoStop}%
\bibitem [{\citenamefont {Fily}\ and\ \citenamefont {Marchetti}(2012)}]{marcetti2012}%
  \BibitemOpen
  \bibfield  {author} {\bibinfo {author} {\bibfnamefont {Y.}~\bibnamefont {Fily}}\ and\ \bibinfo {author} {\bibfnamefont {M.~C.}\ \bibnamefont {Marchetti}},\ }\bibfield  {title} {\bibinfo {title} {Athermal phase separation of self-propelled particles with no alignment},\ }\href {https://doi.org/10.1103/PhysRevLett.108.235702} {\bibfield  {journal} {\bibinfo  {journal} {Phys. Rev. Lett.}\ }\textbf {\bibinfo {volume} {108}},\ \bibinfo {pages} {235702} (\bibinfo {year} {2012})}\BibitemShut {NoStop}%
\bibitem [{\citenamefont {Fily}\ \emph {et~al.}(2014)\citenamefont {Fily}, \citenamefont {Henkes},\ and\ \citenamefont {Marchetti}}]{fily2014freezing}%
  \BibitemOpen
  \bibfield  {author} {\bibinfo {author} {\bibfnamefont {Y.}~\bibnamefont {Fily}}, \bibinfo {author} {\bibfnamefont {S.}~\bibnamefont {Henkes}},\ and\ \bibinfo {author} {\bibfnamefont {M.~C.}\ \bibnamefont {Marchetti}},\ }\bibfield  {title} {\bibinfo {title} {Freezing and phase separation of self-propelled disks},\ }\href@noop {} {\bibfield  {journal} {\bibinfo  {journal} {Soft matter}\ }\textbf {\bibinfo {volume} {10}},\ \bibinfo {pages} {2132} (\bibinfo {year} {2014})}\BibitemShut {NoStop}%
\bibitem [{\citenamefont {Kelkar}\ \emph {et~al.}(2020)\citenamefont {Kelkar}, \citenamefont {Bohec},\ and\ \citenamefont {Charras}}]{kelkar2020mechanics}%
  \BibitemOpen
  \bibfield  {author} {\bibinfo {author} {\bibfnamefont {M.}~\bibnamefont {Kelkar}}, \bibinfo {author} {\bibfnamefont {P.}~\bibnamefont {Bohec}},\ and\ \bibinfo {author} {\bibfnamefont {G.}~\bibnamefont {Charras}},\ }\bibfield  {title} {\bibinfo {title} {Mechanics of the cellular actin cortex: From signalling to shape change},\ }\href@noop {} {\bibfield  {journal} {\bibinfo  {journal} {Current opinion in cell biology}\ }\textbf {\bibinfo {volume} {66}},\ \bibinfo {pages} {69} (\bibinfo {year} {2020})}\BibitemShut {NoStop}%
\bibitem [{\citenamefont {Tse}\ and\ \citenamefont {Engler}(2010)}]{tse2010preparation}%
  \BibitemOpen
  \bibfield  {author} {\bibinfo {author} {\bibfnamefont {J.~R.}\ \bibnamefont {Tse}}\ and\ \bibinfo {author} {\bibfnamefont {A.~J.}\ \bibnamefont {Engler}},\ }\bibfield  {title} {\bibinfo {title} {Preparation of hydrogel substrates with tunable mechanical properties},\ }\href@noop {} {\bibfield  {journal} {\bibinfo  {journal} {Current protocols in cell biology}\ }\textbf {\bibinfo {volume} {47}},\ \bibinfo {pages} {10} (\bibinfo {year} {2010})}\BibitemShut {NoStop}%
\bibitem [{\citenamefont {Lee}\ \emph {et~al.}(2017)\citenamefont {Lee}, \citenamefont {Erisken}, \citenamefont {Iskratsch}, \citenamefont {Sheetz}, \citenamefont {Levine},\ and\ \citenamefont {Lu}}]{lee2017polymer}%
  \BibitemOpen
  \bibfield  {author} {\bibinfo {author} {\bibfnamefont {N.~M.}\ \bibnamefont {Lee}}, \bibinfo {author} {\bibfnamefont {C.}~\bibnamefont {Erisken}}, \bibinfo {author} {\bibfnamefont {T.}~\bibnamefont {Iskratsch}}, \bibinfo {author} {\bibfnamefont {M.}~\bibnamefont {Sheetz}}, \bibinfo {author} {\bibfnamefont {W.~N.}\ \bibnamefont {Levine}},\ and\ \bibinfo {author} {\bibfnamefont {H.~H.}\ \bibnamefont {Lu}},\ }\bibfield  {title} {\bibinfo {title} {Polymer fiber-based models of connective tissue repair and healing},\ }\href@noop {} {\bibfield  {journal} {\bibinfo  {journal} {Biomaterials}\ }\textbf {\bibinfo {volume} {112}},\ \bibinfo {pages} {303} (\bibinfo {year} {2017})}\BibitemShut {NoStop}%
\end{thebibliography}%
\end{document}


\title{Supplementary Information}

\maketitle

\beginsupplement

\begin{enumerate}
    \item Figure on how shape of each cell is taken
    \item Cell area fluctuations and distribution
    \item Shape distribution heatmap vs time
    \item Comparison to AR distributions from previous studies. Voronoi vs regular segmentation. For both simulations and MDCK. Also add fixed $\calA$ results that don't collapse. Make sure that there is an overall message behind showing these distributions.
    \item Section on all simulation details, table all the parameters used. 
    \item Fitting to the best distribution and perhaps comparing different fits. 
    \item Area plasticity results. 
    \item Individual moments against activity. 
    \item Plot showing stress on DPs for static and plastic simulations, demonstrating that plastic simulations have less stress. 
    
\end{enumerate}
\section{Simulation details}
For figure 3 in the main text, we sweep over the following values of the dimensionless persistence time $\tau/\sqrt{m_vl_0^2/\epsilon_l}$: 25, 50, 100, 150, 200, 250, 300, 350, 450, 500, and 1000. For each persistence time, we simulate the following values of the dimensionless driving force $f_0l_0/\epsilon_l$: 0.00005, 0.0001, 0.0002, 0.0004, 0.0008, 0.0016 and 0.0032. $m_v$ is the mass of each vertex, $l_0$ is the rest length between vertices, and $\epsilon_l$ is the energy cost per unit extension of the perimeter springs. 

For figure 4 in the main text, we sweep over the values of $f_0l_0/\epsilon_l$, $\tau/\sqrt{m_vl_0^2/\epsilon_l}$, and dimensionless plasticity parameter $\eta l_0^2/\sqrt{m_v\epsilon_l}$ in the table below. Each row represents the values of the plasticity parameter,  the persistence times, and the driving forces. For example, the first row represents $5\times12\times1=60$ combinations of plasticity, persistence time, and driving force. 

\begin{table}[h]
    \centering
    {
    \begin{tabularx}{\columnwidth}{|X|X|X|}
        \hline
        $\eta l_0^2/\sqrt{m_v\epsilon_l}$ & $\tau/\sqrt{m_vl_0^2/\epsilon_l}$ & $f_0l_0/\epsilon_l$ \\
        \hline
        0.1, 0.01, 5000, 10000, 100000 & 
        25, 50, 100, 150, 200, 250, 300, 350, 400, 450, 500, 1000 & 
        0.0009 \\
        \hline
        1, 10, 100, 200, 1000 & 
        25, 50, 100, 150, 175, 200, 225, 250, 300, 350, 400, 800 & 
        0.0004, 0.0009, 0.0016 \\
        \hline
        400, 600, 800 & 
        50, 150, 800 & 
        0.0004, 0.0009, 0.0016 \\
        \hline
    \end{tabularx}
    }
\end{table}

\section{Image analysis}

Cell images were segmented using Cellpose \cite{stringer2021cellpose, pachitariu2022cellpose}. Frames of time-series phase contrast images were put into Cellpose and hand-annotated to retrieve the final segmentatations used in our analyses. From these segmentations, we retrieve an image mask for each cell. Boundary pixels are then chosen to create a polygon that accurately represents the shape of each cell without overestimating the perimeters of the cells due to the pixelated nature of the images. Roughly evenly spaced boundary pixels on each cell mask are chosen to achieve this. We then use the relative positions of these points to calculate the area, perimeter, shape parameter $\mathcal{A} = \frac{\left(\sum_{i=1}^{Nv} l_{i} \right)^2}{4 \pi a}$, and aspect ratio $AR = \frac{a}{b}$ of the segmented cells (See Figure S1). 

\begin{figure}[h!]
    \includegraphics[width=\columnwidth]{figure-s1-single-cell-shape-draft5.png}
    \caption{Cell Shape from Segmentation: (a) A portion of a segmentation overlaid on a section of a phase contrast image of an MDCK cell monolayer. The central (magenta) cell is the same as that shown in panel (b). (b) $\ge$ 25 roughly evenly spaced boundary pixels of a segmented cell are chosen from the image mask. The perimeter and area of the polygon formed by these points are used to get the shape parameter $\cal{A}$ of the segmented cell. (c) The ellipse with an equivalent second moment as the segmented cell in panel (b) as specified by the regionprops function in Matlab. The ratio of the major \red{a} and minor \blue{b} axes gives us the aspect ratio (AR) of the cell.
    }
    \label{fig:DPM}
\end{figure}


\begin{figure}[tb]

    \includegraphics[width=\columnwidth]{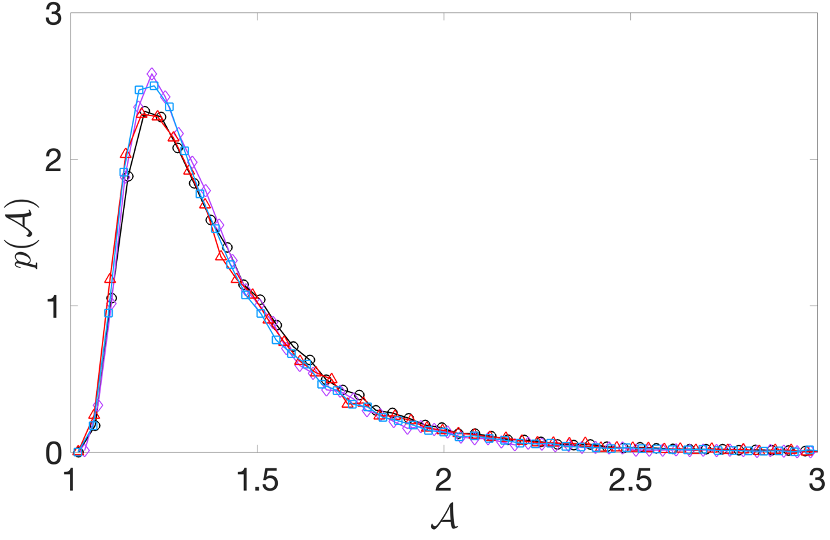}
    \caption{(a) 4 shape distributions taken from the segmentation of 4 different time-lapse image sequences of 4 confluent MDCK cell monolayers. Each shape/color represents analysis from a unique monolayer, each of which was imaged over 3 hours and averaged together for comparison with the simulations presented in the main text of this investigation (Figure 1a). The inset (b) shows the average shape parameter of each monolayer over the course of the imaging. The overall flatness of these curves suggests that the monolayers have reached their steady state shapes over the time range they are analyzed. They also each have similar averages, suggesting agreement with our generalization of our analysis across confluent MDCK monolayers
    }

    \label{fig:prior}
\end{figure}

\section{Comparison to prior work}

Previous studies on the shape distributions found in MDCK cell monolayers have differed from this study in several significant ways. The first is in the segmentation methods used. Previous studies on MDCK cell shapes have largely been conducted using Voronoi tessellations to approximate the outlines of cells in confluent monolayers \cite{atia2018geometric}. This practice limits the cell shapes measured to the unique subset of shapes Voronoi tessellations allow. Voronoi tessellations only allow sharp polygons with about 5 or 6 edges. With the help of advances in cell segmentation technology \cite{stringer2021cellpose, pachitariu2022cellpose}, we are able to capture the rounded curvature of cell shapes in our segmentations, as well as the elongated shapes Voronoi tessellations prohibit. As a result, the segmentation we retrieve from our MDCK cell monolayer images differ notably from those previously reported. We don't see the same collapse onto a single curve when measuring aspect ratio distributions, a significant result in understanding the shapes MDCK cells take in confluent monolayers (See Figure S4). 

\begin{figure}[h!]
    \includegraphics[width=\columnwidth]{figure-s4-lior-figures-vs-segment-vs-voronoi-draft3.png}
    \caption{(a) Aspect ratio (AR) probability distribution functions (PDFs) of MDCK cell monolayers. The black curves are from previously published data \cite{atia2018geometric}. These measurements were retrieved from the ARs of Voronoi tessellations of the cell nuclei. The blue circle curve is from our cell segmentation of MDCK cell monolayer islands using Cellpose \cite{stringer2021cellpose,pachitariu2022cellpose, saraswathibhatla2020spatiotemporal}. The red circle curve is from the ARs of the Voronoi tessellations of the centers of this segmentation. (b) PDFs of a  rescaled measure $x = \frac{AR-1}{\overline{AR}-1}$, where $\overline{AR}$ is the average AR for each distribution respectively.
    }
    \label{fig:AR}
\end{figure}


\section{Fitting the MDCK shape distribution}
\begin{figure}[h!]
    \includegraphics[width=\columnwidth]{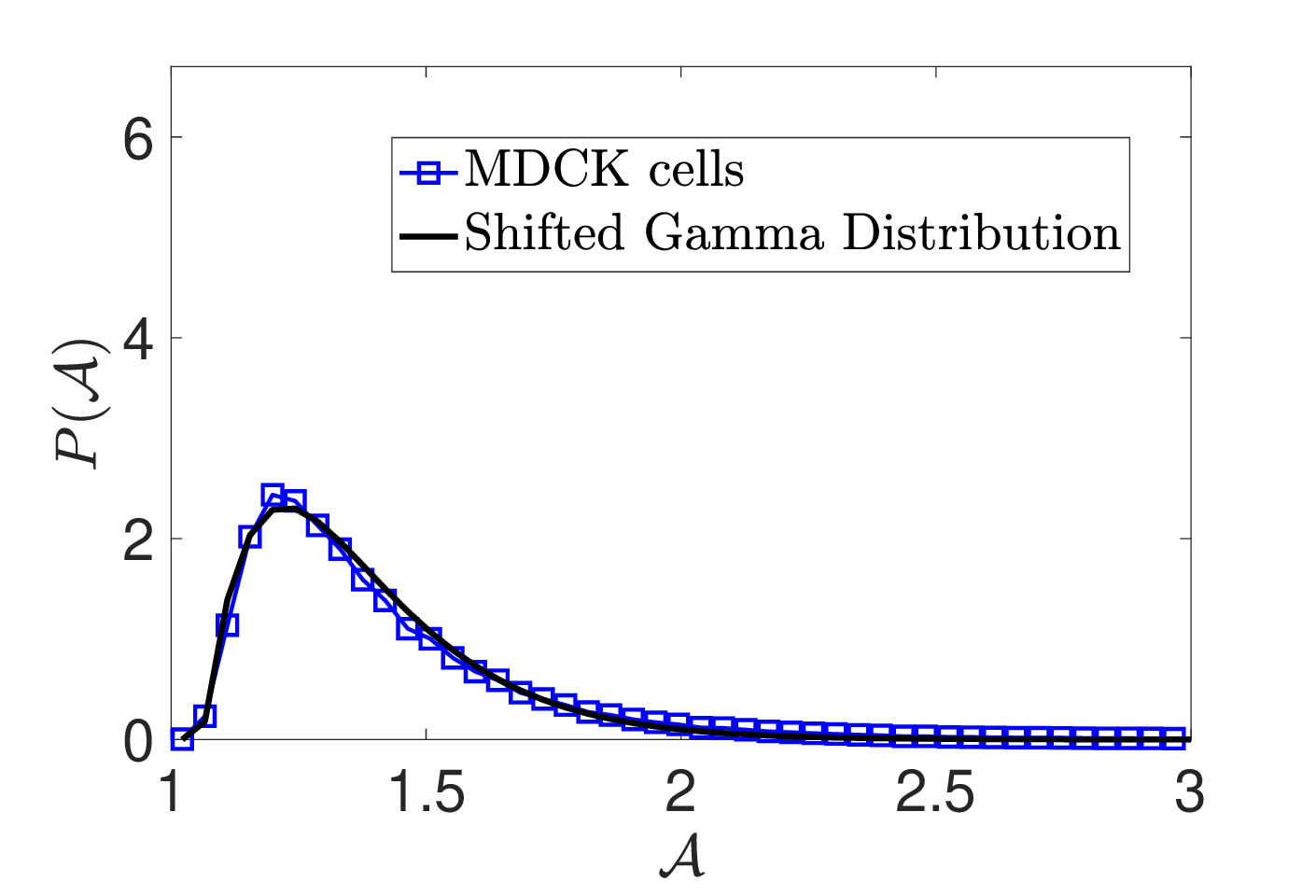}
    \caption{The blue curve shows the segmentation results of the MDCK cells. The red shows the plastic simulation that matches best. Finally, the black curve shows the shifted gamma distribution that best fits the MDCK shape distribution. The fitting parameters are $k=2.03$, $\theta=0.16$, and $s=1.06$.}
    \label{fig:DPM}
\end{figure}


\clearpage
\bibliography{references}